\documentclass[fleqn,usenatbib]{mnras}

\usepackage{newtxtext,newtxmath}

\usepackage[T1]{fontenc}

\DeclareRobustCommand{\VAN}[3]{#2}
\let\VANthebibliography\thebibliography
\def\thebibliography{\DeclareRobustCommand{\VAN}[3]{##3}\VANthebibliography}

\usepackage{graphicx}	
\usepackage{amsmath}	
\usepackage{float}
\usepackage{natbib}
\usepackage{morefloats}
\usepackage{dcolumn}
\usepackage{geometry}
\usepackage{latexsym}
\usepackage{longtable}
\usepackage{lipsum}
\usepackage{gensymb} 
\usepackage{fix-cm}
\usepackage{adjustbox}
\usepackage{lscape}
\usepackage{empheq}
\usepackage{mathrsfs}
\usepackage{textcomp}
\usepackage{hyperref}
\usepackage{multirow}
\usepackage{comment}
\usepackage{booktabs} 
\hypersetup{colorlinks, linkcolor={blue}, citecolor={blue}, urlcolor={blue}} 
\usepackage{xspace}

\usepackage{xcolor}

\newcommand\swift{\textit{Swift}\xspace}

\title[Multi-wavelength studies of ZTF20abbiixp / GRB~200524A]{Exploring the multi-wavelength properties of the high energetic event ZTF20abbiixp/GRB 200524A: from prompt emission to afterglow}

\author[Ghosh et al.]{A. Ghosh$^{1}$\thanks{E-mail: ghosh.ankur1994@gmail.com},
Dimple$^{2,3}$,
K. Misra$^4$,
P. Yu. Minaev$^{5}$,
Y. Yao$^{6}$,
D. A. Kann$^{7}$,
M. Blazek$^{7}$,
A. S. Pozanenko$^{5,8}$,
\newauthor
S. Belkin$^{9}$,
L. Izzo$^{10, 11}$,
H. Kumar$^{12,13}$,
A. de Ugarte Postigo$^{14,15}$,
A. Rossi$^{16}$,
G. C. Anupama$^{17}$,
V. Bhalerao$^{12}$,
\newauthor
D. Bhattacharya$^{18}$,
N. K. Chakradhari$^{19}$,
S. Chandra$^{20}$,
R. Gupta$^{4,21,22}$,
K. M. Jayasurya$^{23}$,
A. Kumar$^{24}$,
\newauthor
B. Kumar$^{25,26}$,
T. S. Kumar$^{4}$,
A. Moskvitin$^{27}$
S. B. Pandey$^{4}$,
A. Omar$^{28}$,
A. R. Rao$^{29}$,
L. Resmi$^{20}$,
\newauthor
V. Rumyantsev$^{30}$,
P. Sanwal$^{4}$,
A.V. Volnova$^{5}$,
A.O. Novichonok$^{31,32}$,
M. Krugov$^{33}$,
S.A. Ehgamberdiev$^{34,35}$,
\newauthor
A.E. Volvach$^{30}$
\\
\\
Affiliations are listed at the end of the paper
}

\date{Accepted XXX. Received YYY; in original form ZZZ}

\pubyear{2026}

\begin{document}
\label{firstpage}
\pagerange{\pageref{firstpage}--\pageref{lastpage}}
\maketitle
\newpage
\begin{abstract}

We conducted a comprehensive multi-wavelength analysis of a high energetic long-duration ZTF20abbiixp / GRB~200524A detected by \textit{Fermi} Gamma Ray Burst Monitor (GBM). Our study combines extended high-energy observations from multiple space-based observatories including \textit{Fermi} with broadband afterglow data spanning X-ray to radio wavelengths, complemented by extensive photometric and spectroscopic follow-up from several ground-based optical facilities worldwide like 3.6-m Devasthal Optical Telescope (DOT). ZTF20abbiixp / GRB~200524A exhibits almost negligible spectral lag, likely arising from the presence of multiple overlapping emission episodes, a property uncommon among long-duration bursts. The burst additionally shows a clear intensity-tracking evolution of the prompt-emission spectral parameters. The broadband afterglow light curve best fits with a broken powerlaw with a break at $10^{5}$ s since the GBM trigger. The electron powerlaw index (p) calculated from the temporal and spectral slopes fail to distinguish between a interstellar medium and a wind environment. Our custom-developed afterglow model fits the panchromatic data well, combining forward shock (FS) and reverse shock (RS) emission. The RS contribution required to fit the early time optical data. The inferred afterglow model parameters suggest that ZTF20abbiixp / GRB~200524A is a high energetic burst expanding into a dense ISM environment, with a relatively large value of the fraction of energy going to accelerating electron and magnetic field ($\epsilon_B$).  
\end{abstract}




\section{Introduction}
\label{introduction}

Gamma-ray bursts (GRBs) are among the most energetic transients in the Universe, releasing an isotropic equivalent energy of $E_{iso} \sim 10^{48} - 10^{55}$ erg in the form of $\gamma$-rays. The two subclasses of GRBs (the long soft and the short hard) are classified based on their $T_{90}$ duration and hardness of the spectra \citep{1993ApJ...413L.101K, 2006ARA&A..44..507W, 2020MNRAS.492.4613O}. Long soft GRBs are associated with the collapse of massive stars \citep{1993ApJ...405..273W}, while short hard GRBs are linked to the merger of compact object binaries (Neutron Star - Neutron Star (NS-NS), Neutron Star - Black Hole (NS-BH)) \citep{1989Natur.340..126E, 2006ApJ...648.1110B, 2017ApJ...850L..21K, 2017ApJ...848L..13A}. 
A subset of long soft GRBs, detected at high energies ($\geq$ 100 MeV) by instruments like the \textit{Fermi} Large Area Telescope (LAT:\citealt{Atwood2009}), shows emission beyond the conventional keV–MeV band. Their isotropic-equivalent energy can reach between $10^{52}$ - $10^{54}$ erg, placing them among the most energetic events \citep{2013ApJS..209...11A}. 
The generic fireball model is based on the blast waves emerging during the formation of the central engine and their interaction within the jet itself or with the ambient medium. Observationally, the GRB emission is divided into two distinct phases - i) The prompt emission phase (short lived, emission in $\gamma$-rays and sometimes in X-rays and optical;  \citealt{1995ARA&A..33..415F}), ii) The afterglow phase (long lived, emission across the electromagnetic spectrum). The modeling supported by the observations in different frequency regimes using ground or space based telescopes, will provide crucial information on the energetics of the burst and physical conditions of the surrounding medium.

The prompt emission mechanism of GRBs was poorly understood before the launch of the \textit{Fermi} satellite and has radically changed since then. Several fundamental questions remain unresolved despite decades of extensive studies of prompt emission, such as the composition of the relativistic jet, the dominant radiation mechanisms, and the characteristic emission radii  \citep{2015PhR...561....1K, 2015AdAst2015E..22P}. The time-integrated spectra of GRBs are predominantly non-thermal in nature and are commonly well explained by the empirical Band function \citep{Band1993}. 
Although the Band function often remains a satisfactory fit for a few bins, alternative physical models, such as a blackbody (BB), a cutoff powerlaw (CPL), or their combinations, including powerlaw plus blackbody (PL+BB) and Band plus blackbody (Band+BB), can also fit some of the temporal bins \citep{2010ApJ...709L.172R, 2011ApJ...730..141Z}. 

The temporal evolution of the peak energy ($E_{\rm p}$) and the low-energy spectral index ($\alpha$) of the Band function has been extensively investigated in the \textit{Fermi} era \citep{2017ApJ...846..137O, 2018A&A...613A..16R, 2019ApJS..242...16L, 2021MNRAS.505.4086G}. Owing to the significant improvement in spectral quality following the launch of \textit{Fermi}, it has been established that approximately two-third of GRBs exhibit a hard-to-soft evolution, in which $E_{\rm p}$ decreases monotonically with time, irrespective of the flux evolution, while rest of the GRBs show intensity tracking pattern where $E_{\rm p}$ varies in tandem with the flux \citep{2012ApJ...756..112L, 2019ApJ...886...20Y}. Despite numerous observational studies, the physical origin of these behaviors remains unclear. Early time-resolved spectroscopy of BATSE data by \citet{1997ApJ...479L..39C} demonstrated that $\alpha$ does not remain constant, but instead evolves with the flux. The evolution of $\alpha$ is more chaotic than the $E_{\rm p}$ evolution. Furthermore, \citet{2018ApJ...869..100U} reported a connection between $E_{\rm p}$ evolution and spectral lag, showing that hard-to-soft evolution is typically associated with positive spectral lag, whereas intensity-tracking is connected with both positive and negative lags.

For the past half-century, relativistic outflows accompanying the core collapse of massive stars have been discovered. A motivation of the Zwicky Transient Facility (ZTF; \citealt{Bellm2019b}; \citealt{Graham2019}) high-cadence survey \citep{Bellm2019a} is to search for optical afterglow from GRBs with too many baryons \citep{Dermer2000}, orphan afterglows \citep{Rhoads1997} from off-axis GRBs, and ordinary afterglows from on-axis GRBs, independent of their high-energy triggers. 
Here we describe the third ZTF-discovered afterglow, ZTF20abbiixp (AT2020kym), associated with GRB\,200524A.

The broadband afterglow emission arises from synchrotron radiation produced as the relativistic jet decelerates through interactions with the ambient medium. The density structure of the medium is influenced by the mass-loss history of the progenitor stars. The broadband afterglow (from radio to VHE $\gamma$-rays) spectrum can be characterized by multiple powerlaw segments separated by characteristic break frequencies normalized to peak flux density \citep{2002ApJ...568..820G, 2014ApJ...781...37P, 2014MNRAS.444.3151V}. The evolution of characteristic frequencies and peak flux depends on the type of medium. As long as GRBs originate from the core-collapse of massive stars, the ambient medium is expected to follow a stellar wind density \citep{2000ApJ...536..195C}. However, \citet{2011A&A...526A..23S} showed that the majority of afterglow light curves are better explained by a constant-density ISM, with only a subset exhibiting signatures of a wind-like profile.

The interaction between the jet and the ambient medium leads to bidirectional shocks: the long-lived forward shock (FS) propagates into the unshocked ambient medium. On the other hand, the reverse shock (RS) is comparatively short-lived downstream to the ejecta and produces optical or Infrared (IR) flare at the very early time (a few tens of s to minutes) or the radio flash at slightly later time \citep{2015AdAst2015E..13G}. The detailed afterglow studies over the past decades have demonstrated that FS emission provides robust constraints on the burst energetics, jet opening angle, and the properties of the ambient medium \citep{2021MNRAS.505.1718J, 2015ApJS..219....9W}, whereas the RS emission offers valuable insights into the jet’s magnetic structure and composition \citep{2015AdAst2015E..13G}. Despite extensive rapid follow-up by \textit{Swift}-UVOT and other optical robotic telescopes, RS signatures were found in a handful of GRBs \citep{2026arXiv260628267G}. \citet{2014ApJ...785...84J} presented a comprehensive afterglow study of 118 GRBs with identified RS signatures in 10 GRBs. The majority of these bursts favor an external-shock origin in a constant-density ISM. 

In this study, we present a detailed multi-wavelength analysis of ZTF20abbiixp / GRB~200524A using a combination of space- and ground-based facilities. The burst was initially detected by \textit{Fermi}-GBM, followed by prompt emission observations from multiple space-based instruments. The afterglow emission was monitored by the \textit{Swift} X-Ray Telescope (XRT) and complemented by extensive follow-up from various ground-based observatories. In section \ref{prompt}, we describe the high energy data acquisition and their temporal and spectral analysis in detail. The afterglow and host galaxy observations using different telescopes are presented in sections \ref{afterglow} and \ref{host}, respectively. In section \ref{results}, we analyze the results. Finally, in section \ref{discussion}, we concluded with a broad discussion of prompt and afterglow emission of ZTF20abbiixp / GRB~200524A. For the analysis, we adopt the standard $\Lambda$CDM cosmology with $H_0= 70 \, \rm km \, s^{-1}\, Mpc^{-1}$, $\Omega_m = 0.27$ and $\Omega_{\Lambda} = 0.73$ \citep{Komatsu2011}. UT times are used throughout the paper. The reported optical photometry is in the AB system, and has been corrected for Galactic extinction $E(B-V)=0.0143$\,mag \citep{Schlafly2011} by adopting the reddening law from \citet{Cardelli1989} with $R_V=3.1$.

\begin{table}
\centering
\caption{Properties of ZTF20abbiixp / GRB~200524A}  
\label{tab:200524A}
 \begin{tabular}{|l|c|}
  \hline
$T_{90}$ duration (s)$^{\dagger}$ &  38.8 $\pm$ 0.9\\
Redshift$^{\ddagger}$ & 1.256\\
Fluence (erg $cm^{-2}$)$^{\dagger}$  &   $(2.08 \pm 0.05) \times 10^{-5}$  \\
$E_{iso}$ (erg) &  (1.80$\pm$ 0.06) $\times 10^{53}$  \\
$E_p$ (keV) &  291 $\pm$ 25  	\\  
$\Gamma_0^{\ast}$  & $503$\\
\hline
\end{tabular}
\newline
\noindent
$^\dagger$ {\it Fermi} - GBM ((7, 850) keV)
\noindent
$^\ddagger$ Spectroscopy
\noindent
$^\ast$ \citet{2010ApJ...725.2209L}
\end{table}

\section{Prompt emission data analysis}
\label{prompt}
\subsection{\textit{Fermi} - GBM temporal analysis}
\label{GBM}

\begin{figure}
	\includegraphics[width=\columnwidth]{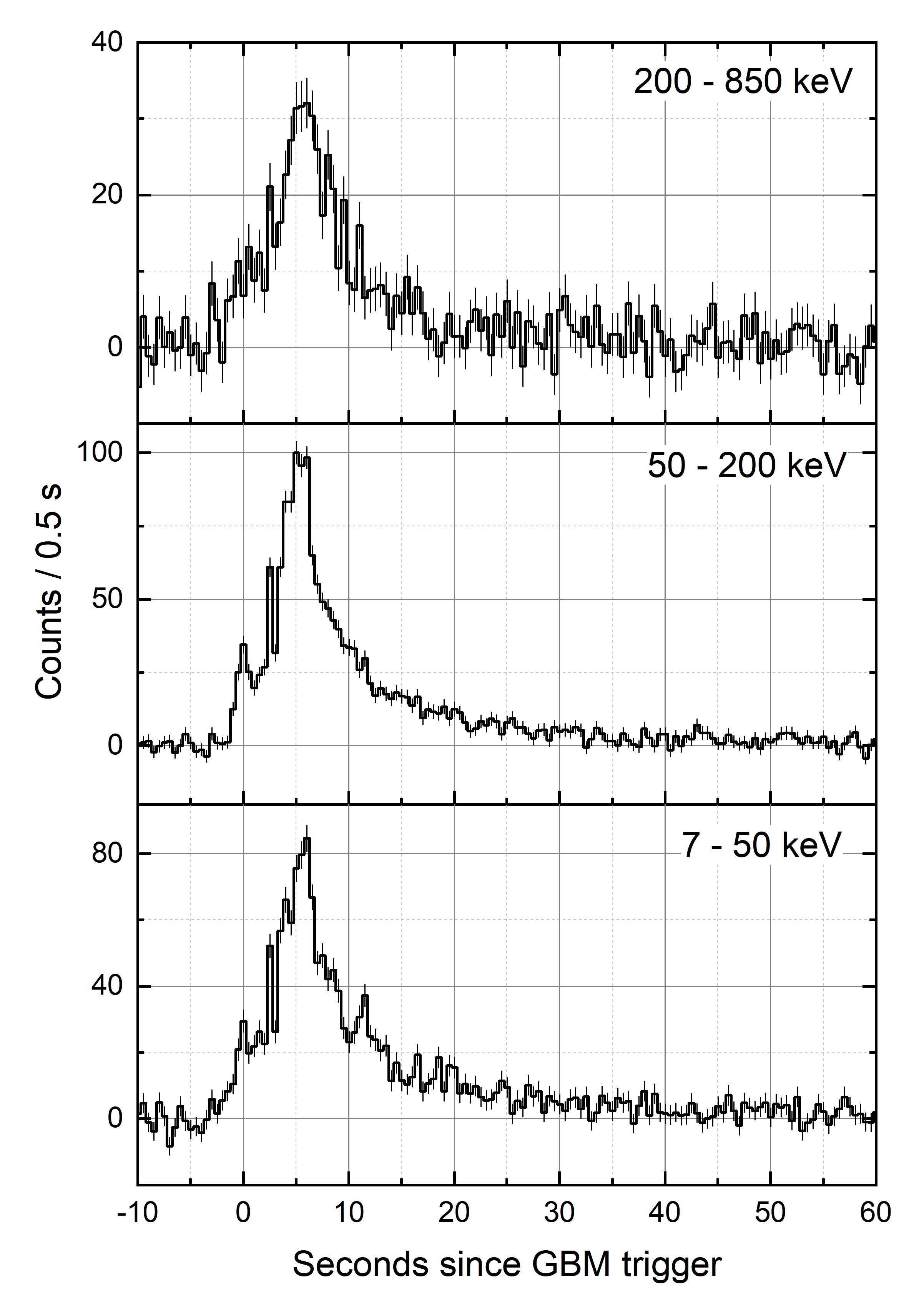}
    \caption{Multichannel light curve of ZTF20abbiixp / GRB~200524A based on {\it Fermi} - GBM data with time resolution of 0.5 s. The boundaries of the energy channels are indicated on the legend.}
    \label{fig:gbm_lcmult}
\end{figure}

ZTF20abbiixp / GRB~200524A was triggered by \textit{Fermi}-GBM (8 keV--40 MeV) at 05:04:00.36 UT on 2020-05-24 \citep{Meegan2009, 2020GCN.27809....1P}. The GBM data were retrieved from the public data archive of \textit{Fermi} satellite \footnote{\url{ftp://legacy.gsfc.nasa.gov/fermi/data}}. Standard data reduction procedures were carried out using packages like \texttt{GTBurst}\footnote{\url{https://fermi.gsfc.nasa.gov/ssc/data/analysis/scitools/gtburst.html}} of \textit{Fermi} Science Tool and \texttt{RMFIT}\footnote{\url{https://fermi.gsfc.nasa.gov/ssc/data/analysis/scitools/rmfit_tutorial.html}} (version 4.3.2). For the analysis, NaI detectors 0, 1, 3, and 5 were selected based on their higher count rates, while the BGO~1 detector was included due to its favorable geometric alignment with the NaI 0-6 detectors. The Time-Tagged Event (TTE) data, with high temporal resolution, were used to construct GBM light curves and to define time intervals based on the burst structure. The burst and background intervals were identified using \texttt{RMFIT}. The linear or cubic polynomial functions were considered to fit the background. The background subtracted light curves were generated in three energy bands (7, 50), (50, 200) and (200, 850) keV as shown in Fig.~\ref{fig:gbm_lcmult}.

\begin{figure}
	\includegraphics[width=\columnwidth]{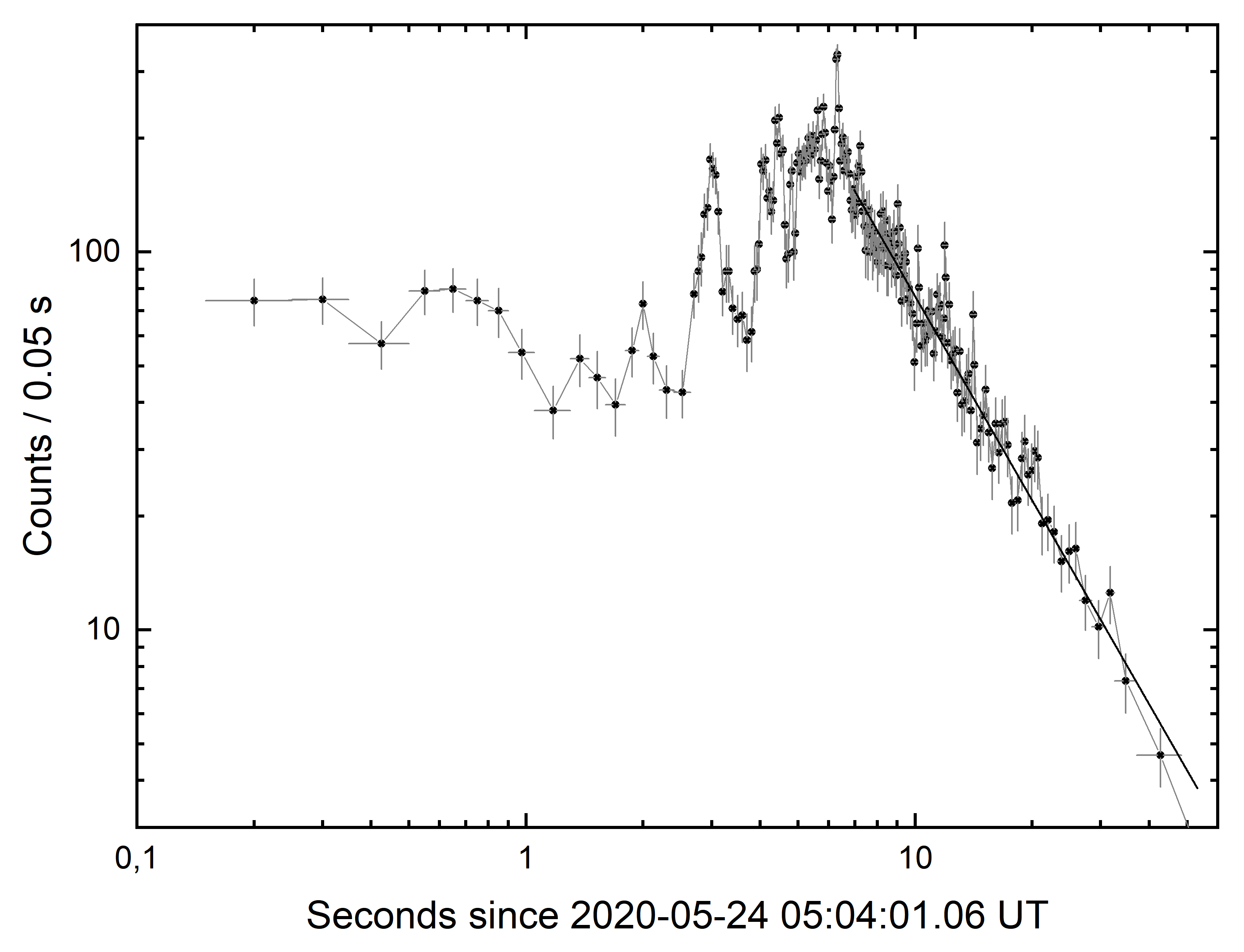}
    \caption{light curve of ZTF20abbiixp / GRB~200524A in energy range (7, 850) keV, based on {\it Fermi} - GBM data. Thick line represents powerlaw fit with index $\alpha$ = -1.79 $\pm$ 0.04.}
    \label{fig:gbm_lclog}
\end{figure}

The duration parameters $ T_\text{90}$ and $T_\text{50}$, the time intervals for which the detector registers 90\% and 50\% of the total number of counts \citep[e.g.][]{kos96}, for ZTF20abbiixp / GRB~200524A are $ T_\text{90} $ = 38.8 $ \pm $ 0.9 s and $T_\text{50}$ = 10.0 $ \pm $ 0.1 s in (7, 850) keV energy range, which is typical for long-soft GRBs. 

Fig.~\ref{fig:gbm_lclog} shows the light curve in the (7, 850) keV energy range in logarithmic scale on both axes. The light curve of ZTF20abbiixp / GRB~200524A exhibits a complex temporal structure composed of several overlapping pulses, with the burst emission evident up to $\simeq$ 50 s. The decay phase of the light curve is well described by a simple powerlaw model with index of $\alpha$ = -1.79 $\pm$ 0.04. Such behavior deviates from the typically observed exponential decay seen in GRB light curves \citep[e.g.][]{nor05,min12,min14}. The resulting fluence in the (7, 850) keV (observer frame bandpass) is $f_{\gamma} = (2.08 \pm 0.03) \times 10^{-5}\,{\rm erg\,cm^{-2}}$. The prompt emission properties of ZTF20abbiixp / GRB~200524A is listed in Table~\ref{tab:200524A}. No evidence for precursor activity is found in the \textit{Fermi}-GBM data. 

\subsection{\textit{Fermi} - GBM spectral analysis}
\label{sec:gbm_spec}

\begin{table*}
\caption{Results of time-integrated and time-resolved spectral analysis of ZTF20abbiixp / GRB~200524A using {\it Fermi}-GBM data.}  
\label{tab:gbm_spec}
\begin{tabular}{cccccc}
\hline
Time interval$^{\dagger}$ & Model$^{\ddagger}$  & $\alpha$ & $\beta$ & $E_\text{p}$ & Flux$^{\ast}$   \\	
(s) & & & & (keV) & ($10^{-7}$ erg cm$^{-2}$ s$^{-1}$)  \\
\hline

\multicolumn{6}{c}{Time-integrated spectral analysis} \\
\hline
(-1, 40) & Band & -0.59 $\pm$ 0.07 & -1.60 $\pm$ 0.02 & 149 $\pm$ 15 & 5.9 $\pm$ 0.1 \\
(-1, 15) & Band & -0.68 $\pm$ 0.05 & -1.73 $\pm$ 0.04 & 201 $\pm$ 19 & 11.5 $\pm$ 0.2 \\
\hline

\multicolumn{6}{c}{Time-resolved spectral analysis} \\
\hline
(-1, 2) & Band & -0.64 $\pm$ 0.20 & -1.53 $\pm$ 0.08 & 192 $\pm$ 81 & 7.3 $\pm$ 0.3 \\
(-2, 4) & Band & -0.41 $\pm$ 0.16 & -1.86 $\pm$ 0.11 & 152 $\pm$ 25 & 11.2 $\pm$ 0.5 \\
(4, 5) & CPL & -0.66 $\pm$ 0.07 & -- & 291 $\pm$ 25 & 19.5 $\pm$ 0.9 \\
(5, 6) & Band & -0.64 $\pm$ 0.07 & -2.3 $\pm$ 0.3 & 258 $\pm$ 29 & 23.8 $\pm$ 0.9 \\
(6, 7) & Band & -0.74 $\pm$ 0.10 & -1.78 $\pm$ 0.10 & 217 $\pm$ 47 & 22.2 $\pm$ 0.8 \\
(7, 9) & Band & -0.74 $\pm$ 0.13 & -1.56 $\pm$ 0.05 & 188 $\pm$ 50 & 14.6 $\pm$ 0.4 \\
(9, 12) & Band & -0.76 $\pm$ 0.13 & -1.68 $\pm$ 0.09 & 181 $\pm$ 45 & 9.0 $\pm$ 0.4 \\
(12, 20) & Band & -0.65 $\pm$ 0.25 & -1.57 $\pm$ 0.05 & 126 $\pm$ 39 & 4.6 $\pm$ 0.2 \\
(20, 40) & PL & -1.34 $\pm$ 0.03 & -- & -- & 2.0 $\pm$ 0.1 \\
\hline

\multicolumn{6}{l}{$^{\dagger}$ Time interval since the {\it Fermi}-GBM trigger.}\\
\multicolumn{6}{l}{$^{\ddagger}$ Best-fitting spectral model.}\\
\multicolumn{6}{l}{$^{\ast}$ Energy flux in the 10 keV -- 1 MeV range.}\\

\end{tabular}
\end{table*}

High energy spectral analysis is a powerful tool to determine the radiation mechanism associated with the prompt emission mechanism of GRBs. \texttt{RMfit v4.3.2} software package was extensively employed to generate and model the time-integrated as well as the time-resolved spectra of ZTF20abbiixp / GRB~200524A. The spectral analysis was carried out according to the method outlined by \citet{gruber14}. Data from detectors with higher count rates and detector source angles $\leq 60^{\circ}$ (NaI\_00, NaI\_01, NaI\_03, BGO\_00) were used for spectral analysis. We considered four spectral models: powerlaw (PL), powerlaw with exponential cutoff (CPL), black body (BB), and Band function \citep{Band1993} to fit the time integrated and time resolved spectra. Table~\ref{tab:gbm_spec} summarizes the results of the spectral analysis. The energy spectrum of the entire burst (time interval (-1, 40) s) is best described by a Band function with $\alpha$ = -0.59 $\pm$ 0.07, $\beta$ = -1.60 $\pm$ 0.02 and $E_\text{p}$ = 149 $\pm$ 15 keV, indicating a non-thermal nature of the spectra of ZTF20abbiixp / GRB~200524A.

\begin{figure}
	\includegraphics[width=\columnwidth]{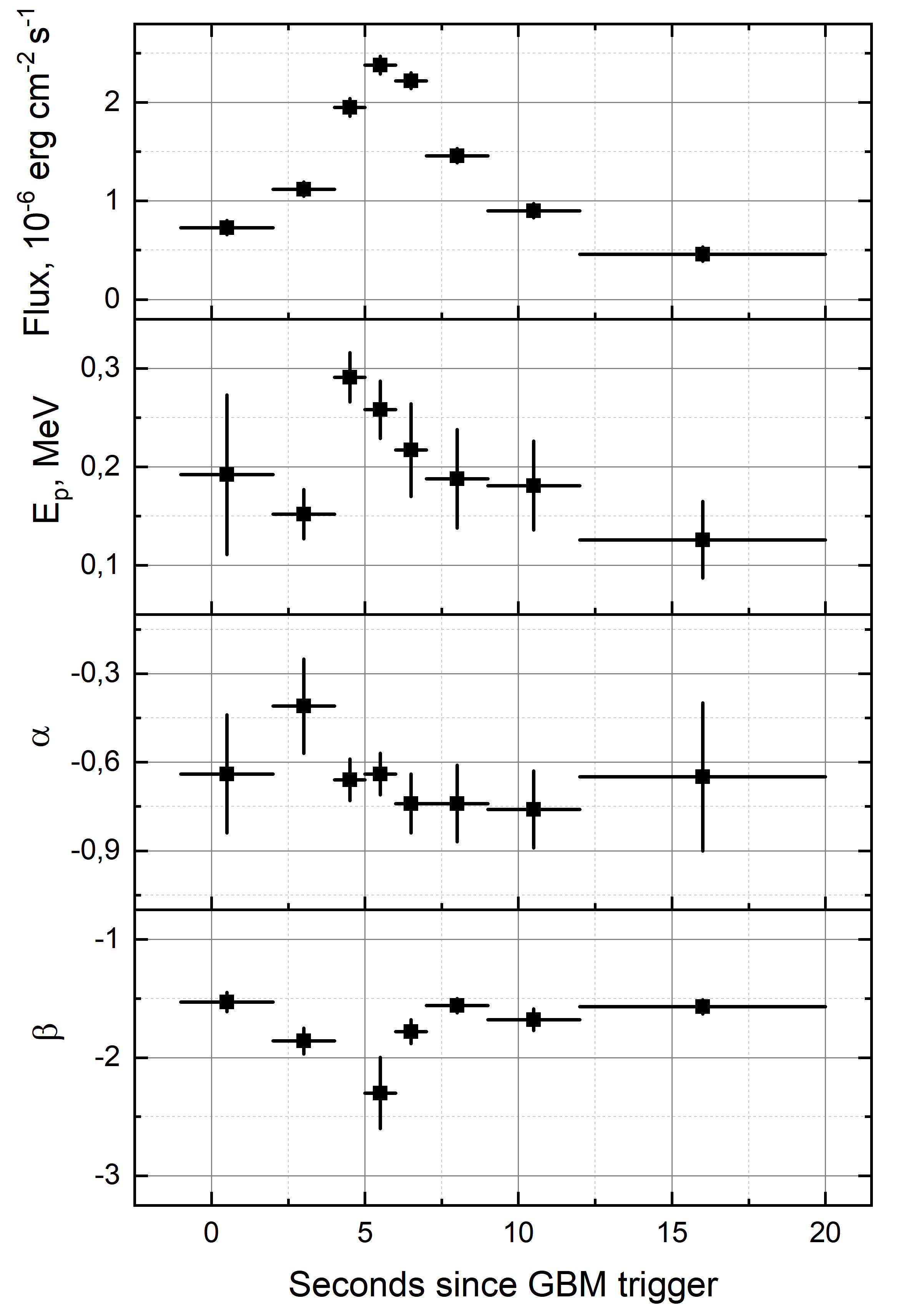}
    \caption{Spectral evolution of ZTF20abbiixp / GRB~200524A main phase based on {\it Fermi} - GBM data (see Table~\ref{tab:gbm_spec}). From top to bottom: the light curve in units of $10^{-7}$ erg cm$^{-2} $ s$^{-1}$, the evolution of the parameter $ E_\text{p} $ in MeV units, the spectral indices $\alpha $ and $ \beta $. The horizontal axis is the time in s relative to the GBM trigger.}
    \label{fig:spevo}
\end{figure}

The time resolved spectral analysis revealed evolution of a few spectral parameters (Fig.~\ref{fig:spevo} and Table~\ref{tab:gbm_spec}). The optimal spectral model for all time bins of the main phase is the Band function, except for the spectrum near the peak of the light curve (time interval (4, 5) s). The time bin near the peak of the spectrum is best described by the CPL model, yielding the highest $E_\text{p}$ = 291 $\pm$ 25 keV. The last bin (time interval (20, 40) s) is best described by a simple powerlaw model, possibly due to poor count statistics. Inclusion of a thermal component failed to provide a satisfactory fit in any of the time bins.

\subsubsection{Spectral parameter correlation}
\label{sp_cor}

\begin{figure}
	\includegraphics[width=\columnwidth]{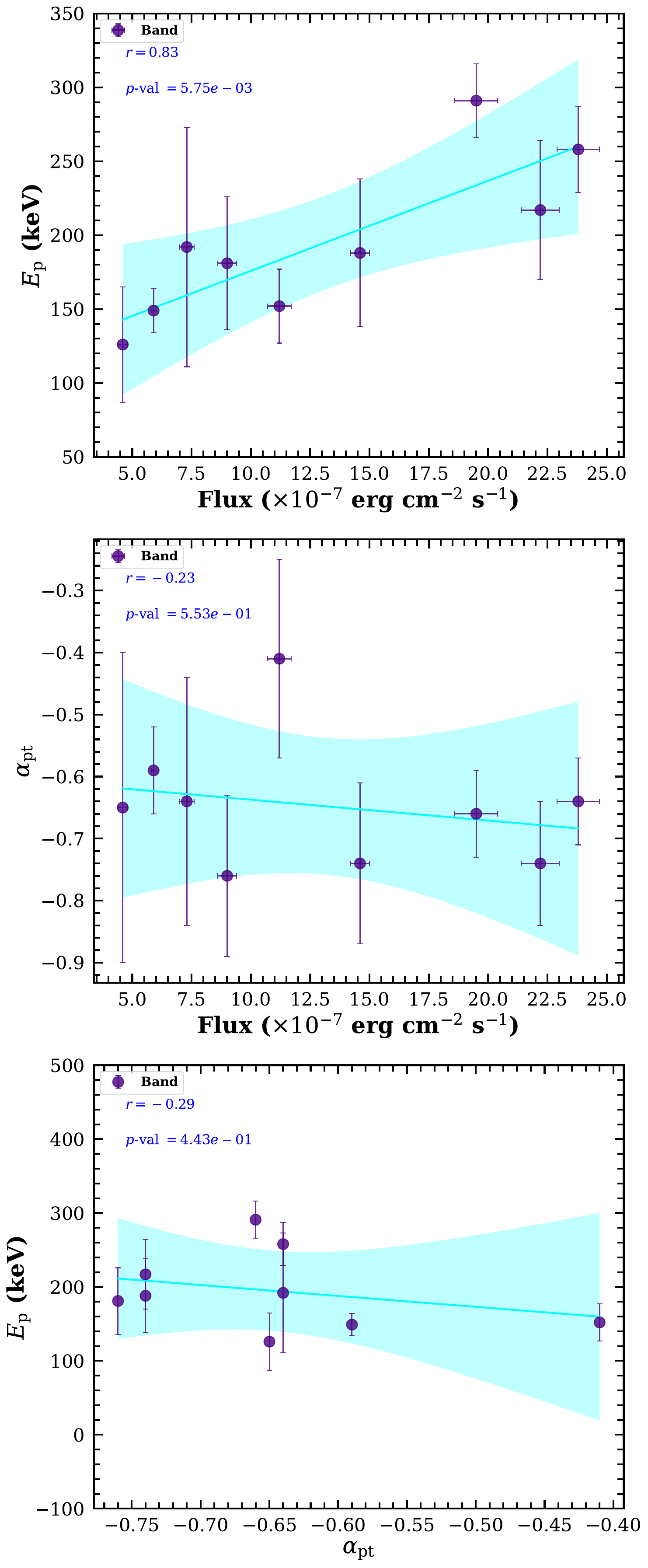}
    \caption{Correlations between spectral model parameters: (a) peak energy vs flux, (b) low-energy spectral index vs flux, (c) peak energy vs low-energy spectral index. Correlations shown in (a), (b), and (c) are obtained using the {\it Fermi} - GBM observations and modelling with the Band function. The best-fit lines are shown in solid indigo, and the indigo-shaded region shows the 3-$\sigma$ confidence interval for the correlations.}
    \label{fig:sp_cor}
\end{figure}

The correlation between three key spectral parameters (flux, $E_{\rm p}$, and $\alpha_{\rm pt}$) obtained from time-resolved spectral analysis of ZTF20abbiixp / GRB~200524A is displayed in Fig.~\ref{fig:sp_cor}. The correlation between the parameters plays a pivotal role in revealing the inter-dependence between the prompt emission spectral properties. The flux for each time interval of the time resolved spectrum was calculated using the Band function, for the energy range between 10 keV to 1 MeV. Similarly, the $\alpha_{\rm pt}$ and $E_{\rm p}$ values were obtained for the same time intervals from the Band function fitting, reported in columns 3 and 5 of Table \ref{tab:gbm_spec}, respectively. 

The relation between the flux and $E_{\mathrm{p}}$ was established in the literature by \citet{1983Natur.306..451G, 2019MNRAS.485.1262B} through the time resolved spectral analysis. Previous studies with \textit{Fermi}-GBM data by \citet{2010A&A...511A..43G, 2019ApJ...886...20Y, 2019ApJ...884..109L, 2021MNRAS.505.4086G} revealed that most of the GRBs follow the non-monotonic trend that is characterized by the broken powerlaw behavior with a turnover at the spectral peak. Only a handful of GRBs show a monotonic trend where the relation follows a single powerlaw, either increasing or decreasing. The flux and $E_{\mathrm{p}}$ for ZTF20abbiixp / GRB~200524A evolve monotonically with a positive slope of 6.09, shown in the top panel of Fig. \ref{fig:sp_cor}. We found a strong correlation between flux and $E_{\rm p}$ with a Pearson coefficient (r) of 0.82 and a significance of correlation (p) of $5.74 \times 10^{-3}$, respectively. To further assess the spectral evolution of the burst, we examined two additional correlations: flux - $\alpha_{\rm pt}$ and $E_{\rm p}$ - $\alpha_{\rm pt}$, in order to evaluate the presence of double-tracking behavior. Neither of these correlations was found to be significant, with p-values substantially exceeding the threshold typically associated with strong correlations (p < 0.01). So ZTF20abbiixp / GRB~200524A shows intensity tracking, but the double tracking feature is missing, unlike reported in \citet{2019ApJ...884..109L, 2021MNRAS.505.4086G}.

\subsection{\textit{Fermi} - LAT unbinned likelihood analysis}

\begin{figure}
	\includegraphics[width=\columnwidth]{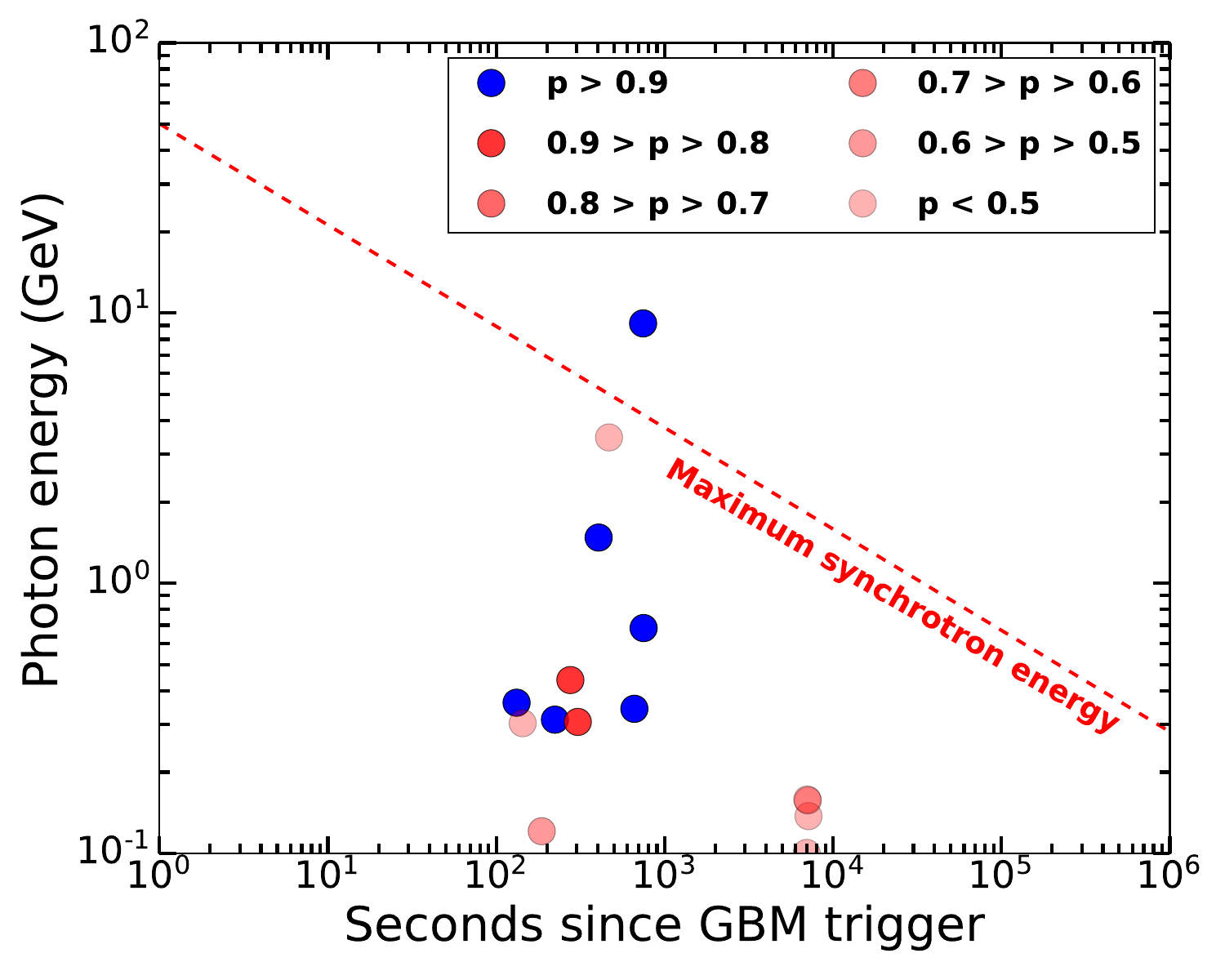}
    \caption{Photons detected by {\it Fermi} - LAT with their respective energies and probabilities of being associated with ZTF20abbiixp / GRB~200524A. The highest-energy photon, with an energy of 9.213 GeV, was detected by the \textit{Fermi}-LAT at $\sim$ 745 s after the GBM trigger.}
    \label{fig:lat photon light curve}
\end{figure}

The high energetic photons ($>100$ MeV) coming from ZTF20abbiixp / GRB~200524A were detected by the \textit{Fermi}-LAT. The emission was located at RA and dec of $\alpha=212.8^{\circ}$, $\delta = 61.0^{\circ}$, respectively, with an error radius of 0.2$^{\circ}$ (90\% containment). The \textit{Fermi}-LAT slewed to the source location at 110 s since the GBM trigger as it was passing through the South Atlantic Anomaly (SAA) \citep{2020GCN.27797....1F}. At the onset of LAT observations, the burst position was located about $25^{\circ}$ from the LAT boresight and lay within $\sim$3$^{\circ}$ of the final GBM position.

The \textit{Fermi}-LAT data of ZTF20abbiixp / GRB~200524A were analyzed by the \texttt{GTBurst} package of \texttt{Fermi Science Tools}, following the standard unbinned likelihood analysis approach. The events in the energy range from 100~MeV to 300~GeV were extracted from a 12$^\circ$ region centered on the burst location to obtain higher count rates. We used a maximum zenith angle of $100^\circ$ to minimize the contamination from cosmic rays and the Earth limb. The time intervals satisfying the good data quality criteria (\texttt{DATA\_QUAL $>$ 0} and \texttt{LAT\_CONFIG = 1}) were implemented. The detector response function \texttt{P8R3\_SOURCE\_V3} was adopted throughout the analysis. The background was modeled considering the galactic diffuse emission (\texttt{gll\_iem\_v07}), extragalactic isotropic component (\texttt{iso\_P8R3\_SOURCE\_V3\_v1}), as well as the contribution from the nearby sources enlisted in the LAT catalog. The \texttt{gtlike} task was used to estimate the source significance using the test statistic (TS) and to constrain the spectral parameters. In addition, the photons associated with ZTF20abbiixp / GRB~200524A were extracted using the tool \texttt{grbsrcprob}.    

The highest energetic photon associated with ZTF20abbiixp / GRB~200524A has an energy $\sim$ 9.8 GeV as shown in Fig.~\ref{fig:lat photon light curve}. The photon flux above 100 MeV obtained from the likelihood analysis is $(2.2\pm0.8)\times 10^{-6}\,{\rm photon\,cm^{-2}\,s^{-1}}$, considering the time bin between $T_0+110$\,s and $T_0+900$\,s.

\subsection{Prompt-emission data analysis using complementary satellite observations}
\label{spi-acs}
ZTF20abbiixp / GRB~200524A was also detected with the anti-coincidence shield (ACS) of the spectrometer SPI on-board the \textit{INTEGRAL} mission. We used a custom developed python routine {\url{http://isdc.unige.ch/$\sim$savchenk/spiacs-online/spiacs-ipnlc.pl}} for the temporal analysis of SPI-ACS data. The light curve of ZTF20abbiixp / GRB~200524A based on \textit{INTEGRAL} - SPI-ACS data is presented in Fig.~\ref{fig:acs_lc} with a time resolution of 0.3 s. It consists of several overlapping pulses with duration parameters of $T_\text{90}$ = 38.3 $\pm$ 0.5 s and $T_\text{50}$ = 10.5 $\pm$ 0.1 s. Based on the characterization of \citet[][]{min10a,min10b}, ZTF20abbiixp / GRB~200524A can be classified as a long soft burst. The total fluence of ZTF20abbiixp / GRB~200524A is $f_{\nu}$ = (11.0 $\pm$ 0.3)$\times 10^4$ counts. Using calibration results, derived in \citet{vig09}, we obtained $f_{\nu}$ $\simeq$ 1.1 $\times$ $10^{-5}$ erg cm$^{-2}$ in energy range (75, 1000) keV. We calculated $F$ $\simeq$ 2.8 $\times$ $10^{-5}$ erg cm$^{-2}$ in (10, 1000) keV range following \citet{poz20}, which is in a good agreement with the value, obtained for ZTF20abbiixp / GRB~200524A using {\it Fermi} - GBM data (see Section~\ref{GBM}). Similar to {\it Fermi} - GBM data, we did not find any precursor emission in \textit{INTEGRAL} - SPI-ACS data following the method described in \citet{min17}.

ZTF20abbiixp / GRB~200524A triggered the Konus (10\,keV--10\,MeV; \citealt{Aptekar1995}) instrument on the \textit{Wind} spacecraft at 2.89 s since the GBM trigger \citep{200524A_KonusDetGCN}. The strongest peak in the multi-peaked light curve lies at $T_0-0.05$\,s. Similar to the \textit{Fermi}-GBM, the time-integrated spectrum of Konus \textit{Wind} is best fit by a Band function with $E_{\rm peak}=215_{-46}^{+48}$\,keV, $\alpha = -0.75_{-0.18}^{+0.25}$, and $\beta = -2.13_{-0.31}^{+0.20}$. The resulting fluence in the 20\,keV--10\,MeV (observer frame bandpass) is $f_{\gamma} = 3.48_{-0.62}^{+0.74}\times 10^{-5}\,{\rm erg\,cm^{-2}}$.

The Cadmium Zinc Telluride Imager (CZTI) on board \textit{AstroSat} \citep{Singh2014} detected 
ZTF20abbiixp / GRB~200524A in the 40--200 keV\,energy range \citep{200524A_AstroSatDetGCN}. The light curve was multi-peaked with a duration of $T_{90}=15.38\pm 0.27$\,s, and the strongest peak was at
$T_0+5.13$\,s. ZTF20abbiixp / GRB~200524A was also detected by the Cesium Iodide (CsI) anticoincidence (Veto) detector in the  100--500\,keV energy range. The light curve was multi-peaked with a duration of $T_{90}=16.89\pm 0.67$\,s, and the strongest peak was at $T_0+3.88$\,s.  Temporal analysis was performed using the standard CZTI\footnote{\url{http://astrosat-ssc.iucaa.in}} pipeline.

\section{Afterglow data analysis}
\label{afterglow}

\subsection{\textit{Swift} observation}
The \textit{Neil Gehrels Swift Observatory} (hereafter \swift; \citealt{Gehrels2004}) initiated a target of opportunity (ToO) observation of ZTF20abbiixp / GRB~200524A following the \textit{Fermi} GBM detection \citep{200524A_SwiftToOGCN}. The X-ray telescope (XRT; 0.3--10\,keV, \citealt{Burrows2005}) collected 5\,ks of data in the photon counting mode between $T_0 + 36.4$\,ks and $T_0+55.7$\,ks and detected four different uncatalogued sources within the uncertainty radius of 4.4 arcsec \citep{200524A_XRTDetGCN}\footnote{XRT full analysis is reported via \url{https://www.swift.ac.uk/ToO_GRBs/00021001/}, and source 6 is the afterglow of the ZTF20abbiixp / GRB~200524A. Standard analysis process of XRT is available at \url{https://www.swift.ac.uk/xrt_products/00021001}.}. Using the automated online tools by \citet{Evans2007, Evans2009}, we model the time-averaged spectrum using an absorbed powerlaw with photon index $\Gamma = 1.8^{+0.9}_{-0.5}$. The XRT light curve (Table~\ref{tab:xrt}) can be fitted with a powerlaw decay ($f_{\rm X} \propto t^{\alpha}$) with a decay index of $\alpha=-0.57_{-0.03}^{+0.05}$. 

In addition to the XRT observations, \swift\ observed ZTF20abbiixp / GRB~200524A with its Ultraviolet/Optical Telescope (UVOT; \citealt{Roming2005}). In the initial follow up, the afterglow falls outside UVOT's field of view \citep{200524A_UVOTMissGCN}. In a further follow up, ZTF20abbiixp / GRB~200524A was detected at $22.06\pm0.12$\,mag in the UVOT white filter \citep{Breeveld2011} in 5.4\,ks exposure taken between $T_0+1.15$\,d and $T_0+1.57$\,d \citep{200524A_UVOTDetGCN}.

\begin{table}
\caption{\swift/XRT observation log of ZTF20abbiixp / GRB~200524A}
\label{tab:xrt}
\centering
 \begin{tabular}{ccc}
  \hline
$\Delta$t & $f_X$  & $f_\nu$     	\\
\addlinespace[0.1cm]
(d) & ($10^{-13}$ erg $s^{-1}$ $cm^{-2}$) & ($10^{-3}$ $\mu$Jy) \\
\hline
\addlinespace[0.15cm]
$0.4931^{+0.1517}_{-0.0722}$ & $2.524\pm 0.490$ & $5.443 \pm 1.057$ \\
\addlinespace[0.15cm]
$1.2586^{+0.1867}_{-0.1066}$ & $1.481\pm 0.395$ & $3.193 \pm 0.852$ \\
\addlinespace[0.15cm]
$1.5598^{+0.0121}_{-0.0087}$ & $<2.568$ & $<5.539$ \\
\addlinespace[0.15cm]
\hline
\end{tabular}
\newline
\newline
\noindent
\noindent
\footnotesize{NOTE: $\Delta t$ is observer-frame time relative to the GBM trigger time $T_0$ given in Section \ref{GBM}. $f_{\rm X}$ is the observed 0.3--10\,keV X-ray flux. $f_{\nu}$ is the flux density at 2\,keV corrected for column density.}
\end{table}

\subsection{ZTF discovery of ZTF20abbiix / GRB~200524A}
ZTF20abbiix / GRB~200524A was discovered by ZTF, which runs on the Palomar Samuel Oschin Schmidt 48-inch (P48) telescope. The ZTF observing system is described in \citet{Dekany2020}. The ZTF data for ZTF20abbiix / GRB~200524A were reduced using the real time photometric pipeline, which employs the image subtraction method outlined in \citet{Masci2019, Zackay2016}. The first real-time alert \citep{Patterson2019} was generated at $T_0+1.802$\,h for an $r$-band detection at $17.35\pm0.04$ mag as part of the ZTF Uniform Depth Survey. During the first epoch observation, the optical counterpart was very bright. A total of four observations between $T_0+1.802$\,h and $T_0+2.287$\,h show continuous fading from $\sim17.4$ to $17.7$\,mag \citep{Ho2020GCN}, and thus passed a set of filters designed to identify afterglow emission in real-time \citep{HoZTF20aajnksq2020}. The detection position is $\alpha = 14^{\mathrm{h}}12^{\mathrm{m}}10.33^{\mathrm{s}}$, $\delta = +60^{\mathrm{d}}54^{\mathrm{m}}19.0^{\mathrm{s}}$ (J2000), which is $0.15^{\circ}$ from the \textit{Fermi} LAT localization of ZTF20abbiix / GRB~200524A, making it a promising optical afterglow candidate. Subsequent follow-up observations carried out by the ZTF collaboration and other telescopes worldwide confirmed the rapid temporal decay of ZTF20abbiix / GRB~200524A, thereby firmly establishing its identification as an afterglow (see section~\ref{optical_observations}) \citep{2022ApJ...938...85H}.
 
\begin{figure*} 
	\includegraphics[width=\textwidth]{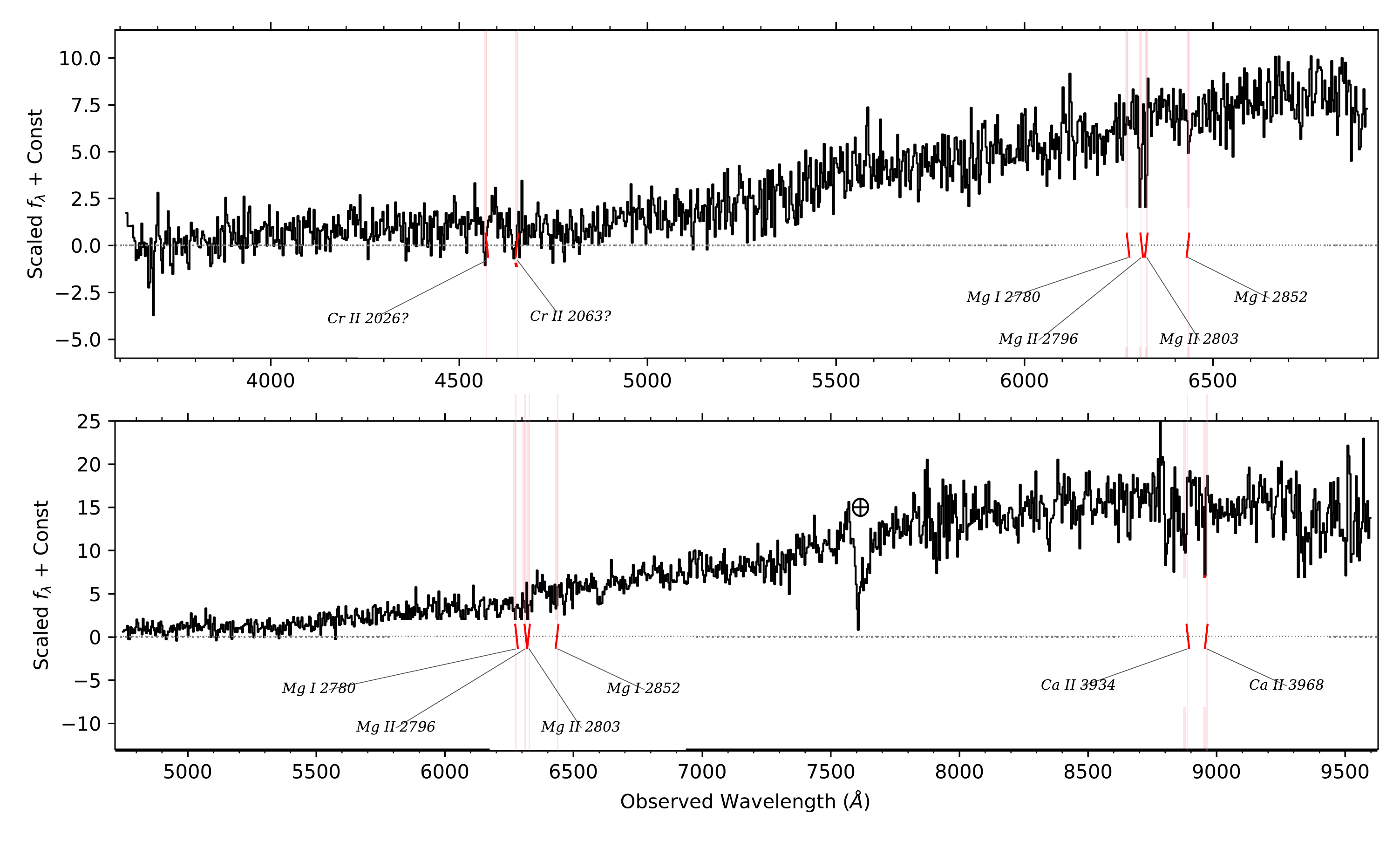}
    \caption{Spectrum of ZTF20abbiixp / GRB~200524A afterglow obtained at $\sim$ $T_0$+27h. The upper and bottom panel show the resulting spectrum in the blue (binned by 3 \AA) and red (binned by 4\AA) grating, respectively. Absorption features of \ion{Mg}{II}, \ion{Cr}{II} and \ion{Ca}{II} at $z=1.256$ are marked.}
    \label{fig:spectroscopy}
\end{figure*}

\subsection{Other optical observations}
\label{optical_observations}

We triggered the GROWTH-India Telescope~(GIT) for optical follow-up of ZTF20abbiixp / GRB~200524A. GIT is located at the Indian Astronomical Observatory~(IAO) in Hanle and is equipped with a 4096$\times$4108 pixel back-illuminated Andor camera. Observations started from 16:12:0.00~UT on 2020-05-24 (11.13 hours since the burst) utilizing g$^\prime$, r$^\prime$ and i$^\prime$ filters. We used 600~s exposure for each frame and reached a median depth of 20.6 mags (5-sigma) in r$^\prime$ filter. The datasets were retrieved in real time using automated scripts. Standard image cleaning was performed using bias and flat-field frames obtained on the same night. Cosmic-ray artifacts were removed using the \textsc{Astro-SCRAPPY} python package \citep{2019ascl.soft07032M}. Source detection was performed with \textsc{SExtractor} \citep{bertin11}, and the sources \citep{bertin11} from the GIT image were cross matched with the PS1 catalogue~\citep{2018AAS...23143601F} using Vizier. Point spread function (PSF) photometry was then performed, and the instrumental magnitudes were calibrated using appropriate photometric zero-points.

Following the GIT observations, the optical counterpart of ZTF20abbiixp / GRB~200524A was further monitored by the 1.3 m Devasthal Fast Optical Telescope (DFOT; \citealt{2012ASInC...4..173S}) and 3.6 m Devasthal Optical Telescope (DOT; \citealt{2018BSRSL..87...29K}) located in Devasthal, India, beginning $\sim$14 hours post \textit{Fermi}-GBM trigger \citep{2020GCN.27806....1K, 2020GCN.27803....1S}. The DOT observations were performed with the ARIES Devasthal Faint Object Spectrograph and Camera (ADFOSC; \citealt{2019arXiv190205857O}) in $g^\prime$, $r^\prime$, $i^\prime$, and $z^\prime$ bands. We continued optical follow-up observations of ZTF20abbiixp / GRB~200524A up to $\sim T_0 + 30$ d post-trigger, using multiple facilities, including the T80 telescope at the Observatorio de Javalambre (OAJ), the Calar Alto 2.2-m telescope, the Kitab-ISON RC-36 telescope, the SAO-RAS Zeiss telescope, the Assy-Turgen AZT-20 telescope, and the CrAO Shajn telescope. Coordinated observations were taken in $ugriz$ filters with the optical imager (IO:O) on the Liverpool Telescope (LT; \citealt{Steele2004}), and in $gri$ with the Spectral Energy Distribution Machine (SEDM; \citealt{Blagorodnova2018}, \citealt{Rigault2019}) on the robotic Palomar 60-inch telescope (P60; \citealt{Cenko2006}) \citep{200524A_LTobs1GCN, 200524A_LTobs2GCN}. Late time deep $g$- and $r$-band images were acquired with the Low Resolution Imaging Spectrograph (LRIS; \citealt{Oke1995}) on the Keck-I telescope on 2020-June-23, with 10\,min of exposure time. LRIS data were processed using \texttt{lpipe} \citep{Perley2019lpipe}, resulted the 5-$\sigma$ upper limits of $g>26.81$\,mag and $r>26.03$\,mag. The details of our optical observations, along with the photometric measurements reported in GCN circulars \citep{200524A_SwiftToOGCN, Ho2020GCN, 2020GCN.27802....1R, 2020GCN.27805....1P, 200524A_UVOTMissGCN, 2020GCN.27813....1O, 2020GCN.27814....1Z, 2020GCN.27820....1B, 2020GCN.27821....1B, 2020TNSAN.112....1H}, are listed in Table~\ref{tab:optical_mag}.
 
We followed the standard procedure to reduce the data obtained from the aforementioned facilities. The preprocessing technique includes bias subtraction, flat fielding and cosmic ray removal from the observed frames. PSF photometry was performed using the \texttt{DAOPHOT} package. The measured instrumental magnitudes were converted to standard magnitudes after correcting for the zero point. 


\subsection{Afterglow spectroscopy and redshift determination}
\label{subsec:specz}

We triggered a long-slit spectrum of ZTF20abbiixp / GRB~200524A between $T_0+26.6$\,h and $T_0+27.8$\,h with GMOS-N under our ToO program GN-2020A-Q-117 (PI: A.~A.~Miller). Observation was performed in the Nod-and-Shuffle mode with a 1$^{\prime\prime}$ slit. We obtained $3\times 550$\,s spectra with the B600 and R400 grating, providing a coverage over the range $\sim$3620--6910\,\AA\ with a resolution of 5\,\AA\ on the blue side, and a coverage of $\sim$4750--9600\,\AA\ with a resolution of 7\,\AA\ on the red side. We did not observe a standard star, and thus no flux calibration was performed. The spectrum was reduced using the \texttt{IRAF} package for GMOS. The resulting spectrum is shown in  Fig.~\ref{fig:spectroscopy}.

On the blue side (upper panel of Fig.~\ref{fig:spectroscopy}), superposed on a relatively weak and noisy continuum, absorption lines of \ion{Mg}{II} $\lambda\lambda2796$,~2803 at $z=1.256$ can be unambiguously identified. We tentatively attribute the two absorption features around $\sim$4600\,\AA\  as \ion{Cr}{II} $\lambda\lambda2026$,~2063 at the same redshift. On the red side (lower panel), \ion{Mg}{II} $\lambda2803$ at $z=1.256$ is detected, although the existence of \ion{Mg}{II} $\lambda2796$ is less clear. Longward of $\sim$7000\,\AA, the spectrum is noisier due to the abundance of sky lines at longer wavelengths. As such, we can not securely identify any absorption system, but instead mark the expected position of \ion{Ca}{II} $\lambda\lambda 3934$,~3969 at $z=1.256$.

In conclusion, the faint continuum at the shortest wavelengths and the imperfect sky subtraction on the longest wavelengths preclude line identifications in those regions. However, the strong absorption lines of \ion{Mg}{II} are similar to other GRB afterglow spectra \citep{Christensen2011}. Therefore, we adopt $z=1.256$ as the redshift of ZTF20abbiix / GRB~200524A.

\subsection{Host galaxy observation}
\label{host}
We observed the field of ZTF20abbiixp / GRB~200524A four months after the burst trigger to detect the host galaxy with the 2.6 m Shajn Telescope (ZTSh). The observations were spread over three nights (2020-09-22, 2020-09-23, 2020-09-24), resulting in a stacked image of exposure time 4.5 hours. The observations were carried out in high humidity conditions. We did not find any source near the afterglow position, which yields a useful limit of R = 24.6 mag. This magnitude is calibrated against USNO-B1.0 field stars.

The field was observed again on 2021-02-05, i.e., eight months after the burst trigger, to search for the host galaxy with the Large Binocular Cameras (LBC) mounted on the Large Binocular Telescope (LBT, Mt. Graham, Arizona, USA). We obtained 20 min (10 dithered exposures of 120 s) in the $r^\prime$ and $z^\prime$ bands, each with poor seeing of 1\farcs6. Unfortunately, we were unable to detect the host galaxy down to the following $3\sigma$ limiting magnitudes $r^\prime>25.4$ and $z^\prime>24.5$ (AB system), calibrated against SDSS field stars. A faint source with magnitude \(r^{\prime}=24.85\pm0.26\) is detected $\delta R\simeq3^{\prime\prime}$ south of the optical afterglow position. The distance between ZTF20abbiixp / GRB~200524A position and the faint object at z = 1.256 is $25.42 \pm 1.44$ kpc. We further
estimate the probability of chance coincidence following the \citet{2010ApJ...722.1946B} formalism,
\[
P_{\rm cc}=1-\exp[-\pi(\delta R)^2\sigma(\leq m)],
\]
where \(\sigma(\leq m)\) is the sky density of galaxies brighter than the
candidate magnitude. For \(\delta R=3^{\prime\prime}\) and
\(r^{\prime}=24.85\pm0.26\), we obtain
\(P_{\rm cc}=0.23^{+0.04}_{-0.04}\). This relatively high value, together with
the large projected offset of $25.42 \pm 1.44$ kpc, indicates that the source cannot be considered as a host galaxy of \ ZTF20abbiixp / GRB~200524A based on positional coincidence alone.

\subsection{Radio Observations}
\label{ref:vla}

After the discovery of ZTF20abbiixp / GRB~200524A, we triggered our ToO program on the Karl G. Jansky Very Large Array (VLA; \citealt{Perley2011}) for fast-rising and luminous transients under the project ID 20A-374 (PI: Anna Y. Q.~Ho). The field was observed for four epochs in X-band (central frequency $\nu = 10$\,GHz). The data reduction steps, including flagging, calibration, and imaging, were performed using the standard VLA pipeline. The flux measurements of the source are tabulated in Table~\ref{tab:vla}. The source was not detected in the last epoch, resulting in a 3-$\sigma$ upper limit. 

We observed the field of ZTF20abbiixp/GRB~200524A with the RT-22 telescope at 36.8 GHz over the interval \(t-T_{0}\simeq1.7\)--4.0 d after the GRB trigger. No radio emission was detected at the optical afterglow position. The resulting upper limits are reported in Table~\ref{tab:vla} and are shown as downward triangles in Fig.~\ref{fig:model}.

\begin{table}
\centering
\caption{Radio observations of ZTF20abbiixp / GRB~200524A} 
\label{tab:vla} 
 \begin{tabular}{lccc}
  \hline
UT Date &  $\Delta$t & Frequency band  & $f_\nu$  	\\
 &  (d) & (GHz)  & ($\mu$Jy)    	\\
\hline
2020-05-28.947 & 4.7365  & 10 & $112 \pm 7$ \\
2020-05-29.307 & 5.0965  & 10 & $184 \pm 5$ \\
2020-06-02.969 & 8.7587  & 10 & $169 \pm 5$ \\
2020-06-28.839 & 35.6282 & 10 & $<24 $ \\
\hline
2020-05-25.968 & 1.757 & 36.8 &  $<2700 $ \\
2020-05-26.980 & 2.769  & 36.8 &  $<2400 $ \\
2020-05-27.981 & 3.769  & 36.8 & $<2700$ \\
\hline
\end{tabular}
\newline
\newline
\noindent
\footnotesize{NOTE: $\Delta t$ is observer-frame time relative to the GBM trigger time $T_0$ given in Section \ref{GBM}.}
\end{table}

\section{Results and Discussion}
\label{results}

\subsection{Prompt emission}

\subsubsection{Spectral lag and lag-luminosity correlation}

\begin{figure}
	\includegraphics[width=\columnwidth]{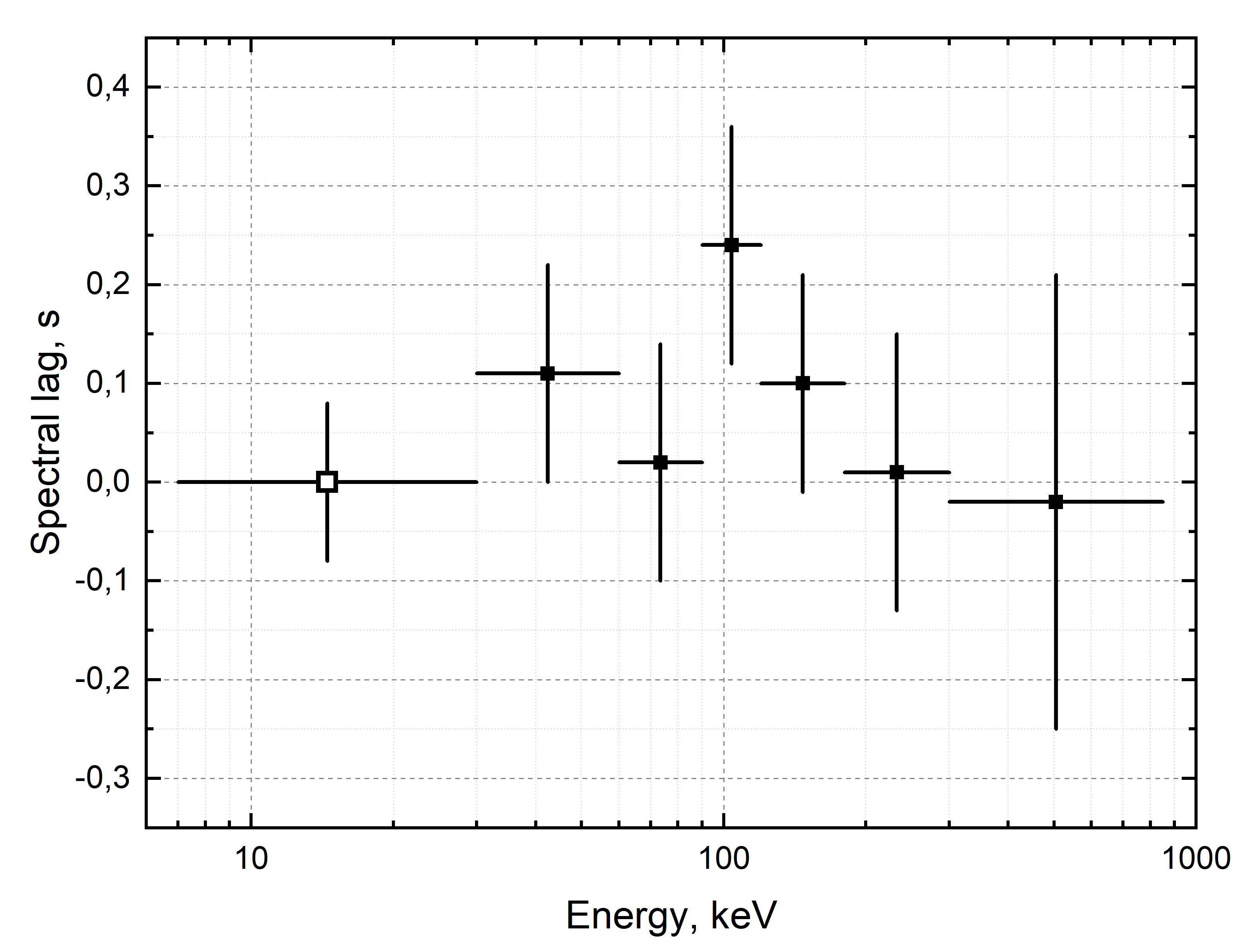}
	\includegraphics[width=\columnwidth]{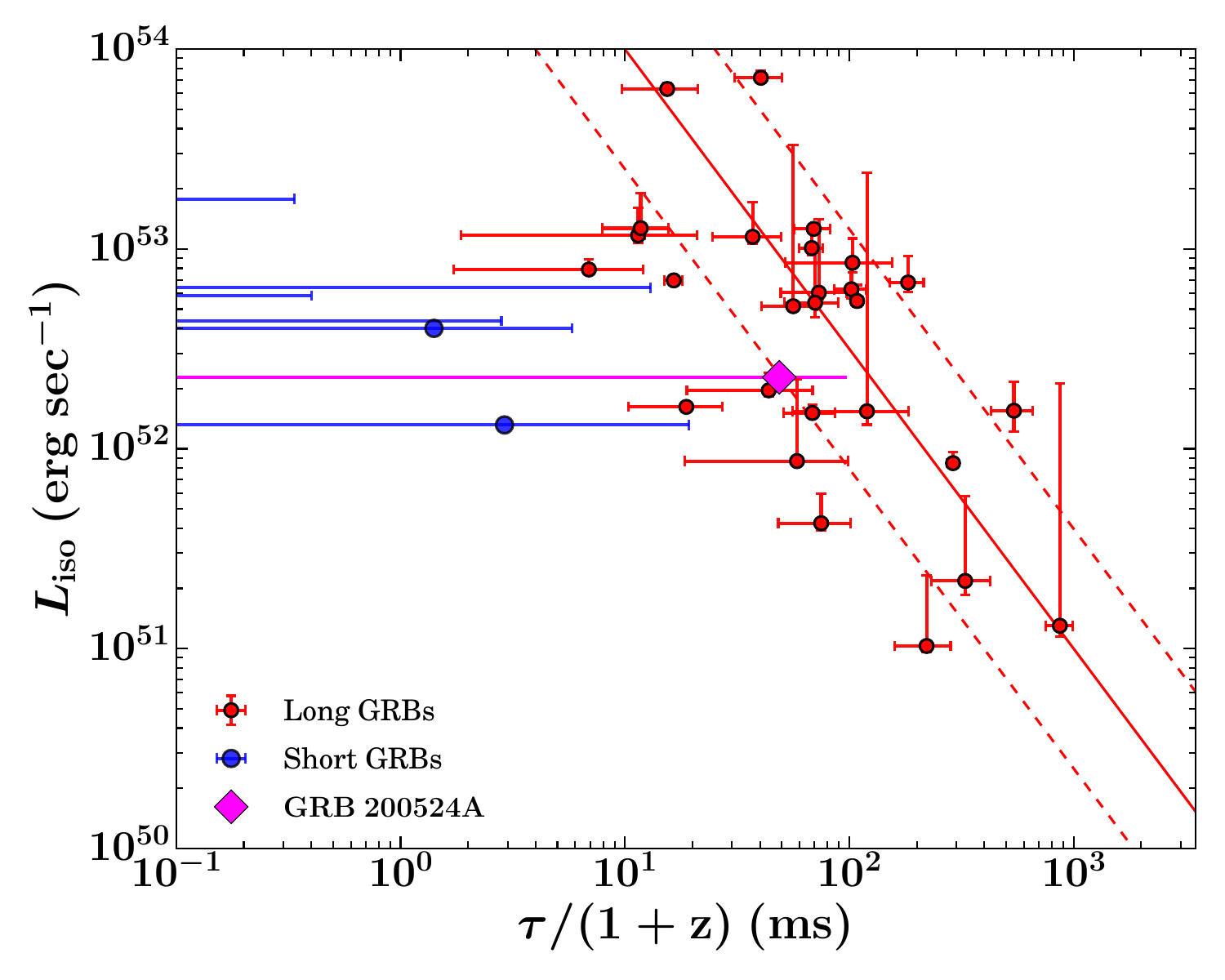}
    \caption{The top panel shows the spectral evolution of ZTF20abbiixp / GRB~200524A based on \textit{Fermi}-GBM data. The horizontal axis -- the energy in units of keV, the vertical axis -- the spectral lag in units of ms relative to the (7, 30) keV channel, shown by an unfilled symbol. The bottom panel shows the spectral lag--luminosity correlation in the rest frame for long soft and short hard GRBs. Long soft and short hard GRBs are shown with red and blue filled circles, respectively. ZTF20abbiixp / GRB~200524A is highlighted with a magenta diamond. The solid red line denotes the best-fitting lag--luminosity anti-correlation for long soft GRBs, and the dashed red lines show the associated 3-$\sigma$ error bars.}
    \label{fig:lags}
\end{figure}

GRBs are characterized by spectral evolution, which can be measured as a relative shift (lag) of the light curves in different energy bands. The lag is considered positive if the hard emission precedes the soft one, and it can be significant (up to a few s) for long soft bursts.

To investigate the spectral lag for ZTF20abbiixp / GRB~200524A, we use the cross-correlation method described in \citet{min12,min14}. The light curves are constructed in seven different energy channels covering the range (7, 850) keV. The softest (7, 30) keV energy channel is chosen as the reference channel, relative to which the cross-correlation of the remaining channels is carried out. The results of the cross-correlation analysis are presented in the upper panel of Fig.~\ref{fig:lags}. Although the burst is expected to have a significant lag, as it represents a long soft population, it demonstrates the absence of a significant spectral lag across the entire energy range (7, 850) keV. Almost negligible spectral lag can be explained by the superposition effect of a large number of overlapping pulses (see Fig.~\ref{fig:gbm_lclog}), each of which has unique spectral-temporal properties \citep{min14}. 

We investigated the position of ZTF20abbiixp / GRB~200524A in the spectral lag - bolometric luminosity correlation. This correlation plot was constructed, incorporating data from both long soft and short hard GRBs as reported by \citet{2000ApJ...534..248N}, \citet{2002ApJ...579..386N}, \citet{2006Natur.444.1044G}, and \citet{2012MNRAS.419..614U}. A clear anti-correlation was found between these two parameters, with a slope of $-1.2 \pm 0.2$. The spectral lag of ZTF20abbiixp / GRB~200524A was computed between the 10–30 keV and 50–100 keV energy bands to ensure consistency with the methodology adopted in \citet{2012MNRAS.419..614U}. With a measured spectral lag of $250 \pm 100$ ms and the isotropic luminosity $L_{\rm iso} = 2.1 \times 10^{52}$ erg/s, ZTF20abbiixp / GRB~200524A is lying close to the boundary of 2$\sigma$ confidence region of the correlation (see the bottom panel of Fig.~\ref{fig:lags}. Such a small spectral lag observed in long soft GRBs is unusual.

\subsubsection{The $E_\text{p,i}$ -- $E_\text{iso}$ correlation (Amati correlation)}

\begin{figure}
	\includegraphics[width=\columnwidth]{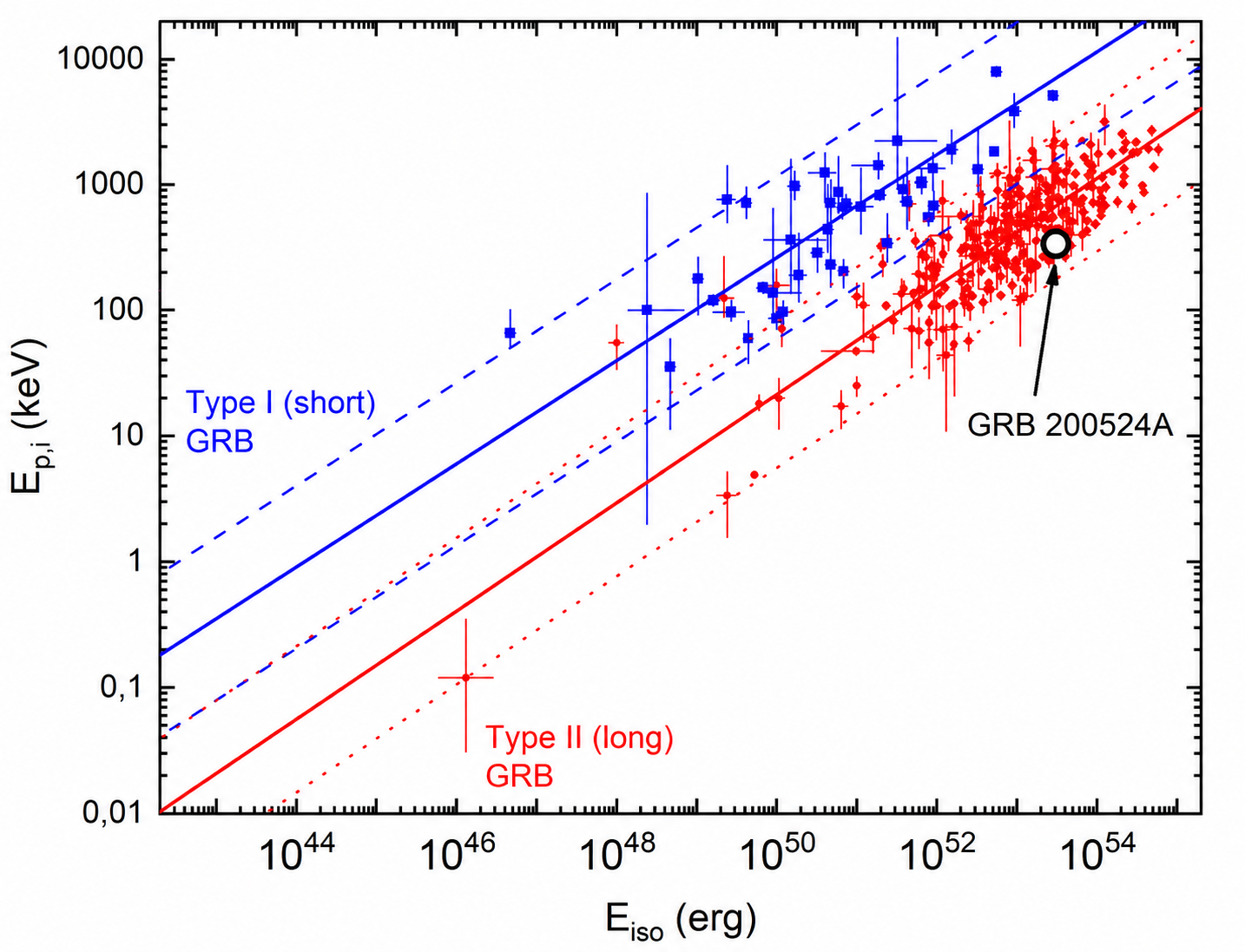}
    \caption{The $ E_\text{p,i} $ -- $ E_\text{iso} $ correlation of for short hard (blue squares) and long soft (red circles) GRBs with the approximation results, including 2$\sigma_\text{cor}$ correlation regions, shown by corresponding colors. The position of ZTF20abbiixp / GRB~200524A is shown by an unfilled black circle, which is typical for long soft GRBs.}
    \label{fig:amati}
\end{figure}

It was shown in \citet{min20b,min20} that the correlation between the isotropic equivalent of the total energy emitted in the $\gamma$-ray range, $ E_\text{iso} $ and the position of the maximum in the energy spectrum $ \nu F_{\nu} $ in the source frame, $ E_\text{p,i} $ (equation~\ref{eq:am}), can be effectively used to classify the GRBs and estimate their redshift. This is facilitated by the observational fact that the correlation for various types of investigated $\gamma$-ray transients is described by a powerlaw with a similar index of $ a \simeq 0.4 $, while the correlation regions of short hard and long soft GRBs are well distinguished. Fig.~\ref{fig:amati} shows the correlation for the sample of 316 GRBs from \citet{min20} with minor corrections from \citet{min21}.

\begin{equation}
    \lg\Big(\frac{E_\text{p,i}}{100~{\text{keV}}}\Big) = a\lg\Big(\frac{E_\text{iso}}{10^{51}~{\text{erg}}}\Big) + b.
	\label{eq:am}
\end{equation}

Adopting the redshift of ZTF20abbiixp / GRB~200524A $z = 1.256$ and using the spectral results obtained from the {\it Fermi}-GBM analysis, we calculate the isotropic-equivalent energy of $E_{\rm iso} = (30.4 \pm 1.6)\times10^{52}$ erg and a rest-frame peak energy of $E_{\rm p,i} = 336 \pm 34$ keV. The position of the burst in the $E_{\rm p,i}$–$E_{\rm iso}$ (Amati) plane is presented in Fig.~\ref{fig:amati}. The property of the burst is consistent with the long soft GRBs, but it is located toward the edge of the 3$\sigma$ confidence region. 

\subsubsection{The $T_\text{90,i}$ -- $EH$ diagram}

\begin{figure}
	\includegraphics[width=\columnwidth]{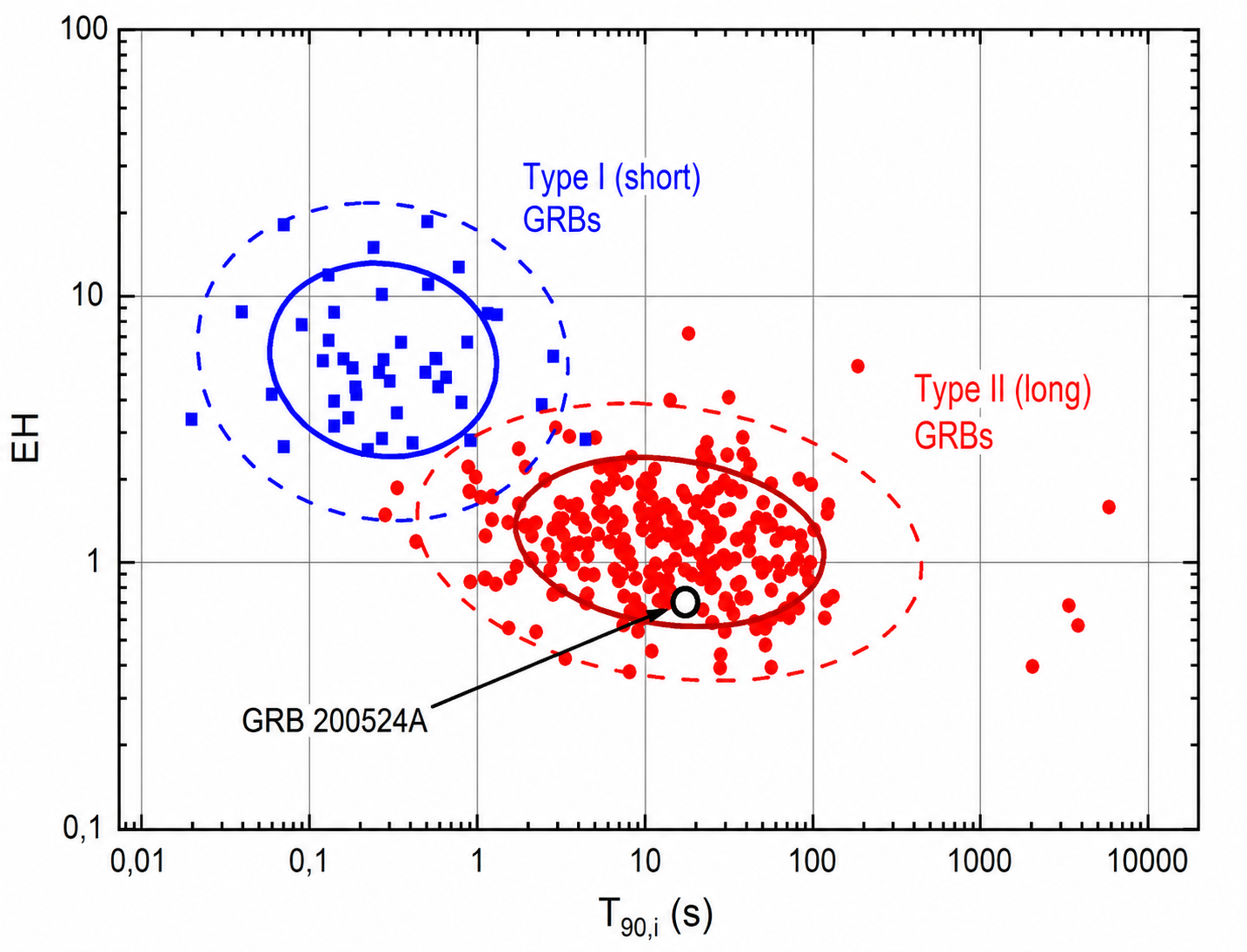}
    \caption{The $ T_\text{90,i} $ -- $ EH $ diagram for short hard (blue squares) and long soft (red circles) GRBs with corresponding cluster analysis results, 1$\sigma_\text{cor} $ and 2$\sigma_\text{cor} $ cluster regions are shown by bold solid and thin dashed curves of the corresponding colors. The position of GRB~200524A is shown by an unfilled black circle.}
    \label{fig:ehd}
\end{figure}

To solve the problem of classification of GRBs, another method was proposed in \citet{min20}, which uses, in addition to the $ E_\text{p,i} $ -- $ E_\text{iso} $ correlation features, the bimodality of the distribution of GRBs in duration in source frame  $ T_\text{90,i}$. For this purpose, the $EH$ parameter (equation~\ref{eq:EH}) was introduced, which characterizes the position of the GRB on the $ E_\text{p,i} $ -- $ E_\text{iso} $ diagram. 

\begin{equation}
    EH = \frac{(E_\text{p,i} / 100~\text{keV})}{ (E_\text{iso} / 10^{51}~\text{erg})^{~0.4}}.
	\label{eq:EH}
\end{equation}

Fig.~\ref{fig:ehd} shows the $ T_\text{90,i} $ -- $ EH $ diagram for the same sample of events, used to construct the $ E_\text{p,i} $ -- $ E_\text{iso} $ diagram. Short hard GRBs, in comparison with long soft GRBs, have a harder spectrum (in terms of $ E_\text{p,i} $ values) with a lower value of total energy $ E_\text{iso} $ and, as a consequence, a larger value of the parameter $ EH$, and also have a shorter duration $ T_\text{90,i}$. 

The corresponding value of the $EH$ parameter is equal to 0.34, while the burst duration in the rest frame is $T_\text{90,i} = 17.2$ s. The position of ZTF20abbiixp / GRB~200524A in the $ T_\text{90,i} $ -- $ EH $ diagram is shown in Fig.~\ref{fig:ehd}, placing the burst into a cluster of long soft bursts, confirming initial classification. 

\begin{figure*}
	\centering
	\includegraphics[width=\textwidth]{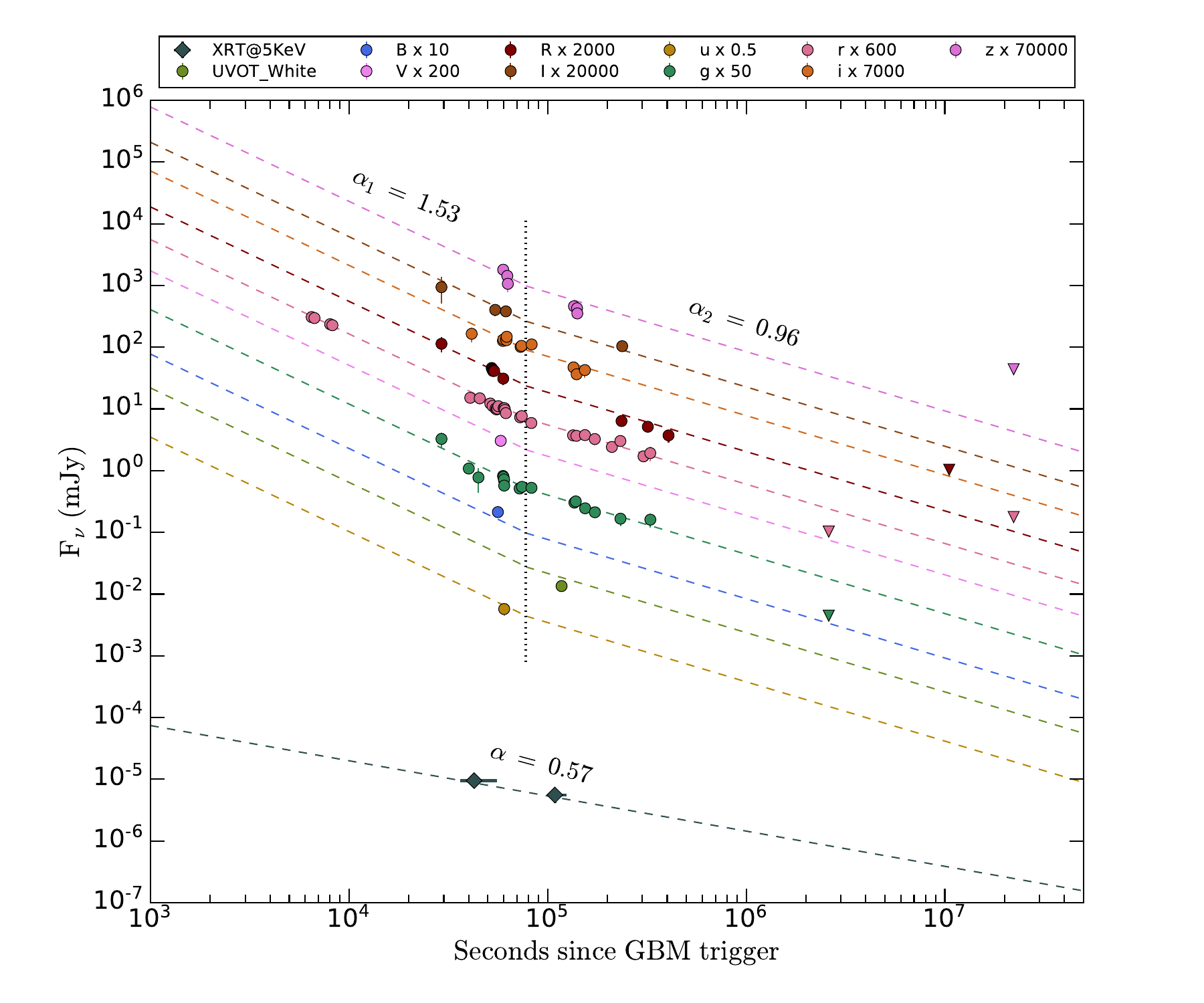}
	\caption{Multi-wavelength afterglow light curves of ZTF20abbiixp / GRB~200524A, including the X-ray flux density at 5 keV (black diamonds) and optical observations in different filters (Coloured symbols). The dashed lines represent the best-fitting powerlaw components in different bands. The dotted vertical line marks the transition time $t_{\rm flat}$} of the light curve.
	\label{fig:lcmult}
\end{figure*}

\subsection{Multi band afterglow light curves of ZTF20abbiixp / GRB~200524A}

\subsubsection{Preliminary Considerations}

The panchromatic light curve of ZTF20abbiixp / GRB~200524A, spanning X-ray to radio frequencies, was constructed by converting calibrated magnitudes into flux densities, as listed in Table~\ref{tab:optical_mag}. The optical data was complemented by X-ray observations from the \textit{Swift}-XRT. 

The optical flux decay seems to exhibit a steep-to-shallow transition (see Fig.~\ref{fig:lcmult}). Therefore, we fit the $ugriz$ light curves with a broken powerlaw $f_\nu(t) \propto t^{-\alpha}$, where $\alpha$ transitions from $\alpha_1$ to $\alpha_2$ at time $t_{\rm flat}$. Non-detections are excluded from the fitting. We utilized the \textsc{emcee} \citep{Foreman-Mackey2013} to explore the parameter space, using 100 walkers with a chain length of 5000 iterations. We ignored the first 1000 iterations to allow the fit to ``burn in'', and then the solution converged. We estimated 68\% confidence intervals using \texttt{corner} \citep{Foreman-Mackey2016}, which gives $\alpha_1 = 1.53^{+0.02}_{-0.01}$, $\alpha_2 = 0.96_{-0.17}^{+0.07}$, $\beta = 0.91_{-0.12}^{+0.11}$, and $t_{\rm flat}=0.81_{-0.10}^{+0.31}$\,d. This indicates that the decay rate of the optical afterglow has slowed down approximately 1 d after the burst. Modulation of the light curve can be caused by interstellar scintillation (ISS; \citet{2013A&A...552A..93H}). Using the closure relations \citep{2018ApJ...866..162G} and the derived values of $\alpha$ and $\beta$, the estimated value of electron powerlaw index (p) ranges between 1.6 to 3.0. The closure relations do not conclusively favor a specific circumburst environment, underscoring the necessity for multi-wavelength modeling.

\subsubsection{Multi-wavelength modeling}

\begin{figure*}
	\includegraphics[width=\textwidth]{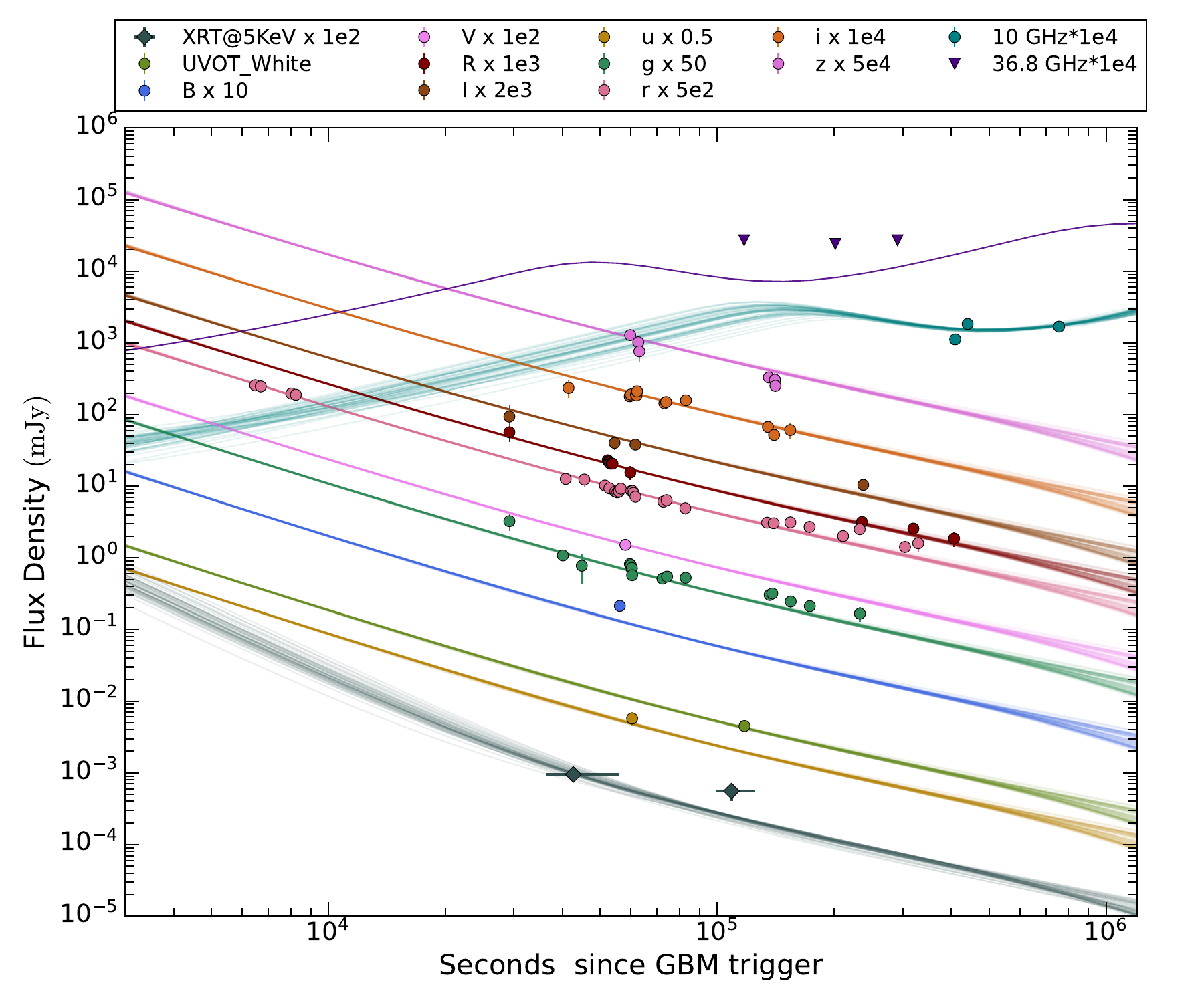}
    \caption{The forest plot obtained from the multi-wavelength modeling of ZTF20abbiixp / GRB~200524A from X-ray to radio bands using our custom-developed FS-RS model. Similar color and marker conventions were followed as in Fig.~\ref{fig:lcmult}. The radio upper limits are shown with downward triangles.}
    \label{fig:model}
\end{figure*}

Based on the multi-wavelength (X-ray to radio) observations from 0.075 to 8.75 d, we constructed well-sampled light curves ZTF20abbiixp / GRB~200524A. The afterglow data were modeled within the framework of the standard relativistic blast-wave scenario, incorporating the combined effects of FS and RS, following the detailed prescription by \citet{2000ApJ...543...66P, 2002ApJ...568..820G, 2013NewAR..57..141G, 2015PhR...561....1K}. The expressions for the characteristic synchrotron break frequencies and peak flux densities in both the FS and RS components are presented in \citet{2026arXiv260323359G}. The cut-off frequency ($\nu_{\rm cut}$) and RS crossing time ($t_{\rm X}$) was also included in the model, as described by \citet{2000ApJ...545..807K}. The adopted model also accounts for the jet break and the transition to the deep Newtonian phase.  We did not include synchrotron self-Compton (SSC) in our model, as no very high energy (VHE) observations of this burst have been reported. The outflow was assumed to be initially quasi-spherical and decelerating as it propagated into a circumburst medium characterized by a density profile of the form n(R) = $(A/m_p)R^{-k}$. Two different density profiles have been explored i) a uniform density ISM (k=0) and ii) a wind-like profile (k=2).  

We employed the Bayesian parameter estimation technique \textsc{PyMultiNest} \citep{2014A&A...564A.125B}, which is based on the nested-sampling Monte Carlo algorithm \textsc{MultiNest} \citep{2009MNRAS.398.1601F}, to model the multi-wavelength dataset and determine the posterior distributions of the physical parameters. A set of ten model parameters (the jet opening angle ($\theta_j$), isotropic equivalent energy ($E_{\rm k,iso}$), the ambient medium density ($n_0$), electron powerlaw index (p), the fraction of energy going to accelerating electron and magnetic field ($\epsilon_e$ and $\epsilon_B$) for FS and RS, and RS crossing time ($t_{\rm X}$)) has been considered. Broad prior ranges, consistent with values commonly inferred for GRBs, were adopted for all parameters. Nested sampling was performed for 10,000 iterations. The adopted priors and their corresponding best fit parameter values are tabulated in Table~\ref{Tab:bestfit_fsrs}. The early-time light curves deviate from the FS model. Incorporating an RS component resolves the discrepancy and provides a satisfactory fit. The forest plot and posterior distributions of ZTF20abbiixp / GRB~200524A, obtained from the best-fitting FS+RS model parameters, are shown in Figs~\ref{fig:model} and \ref{fig:corner}, respectively.

\begin{table}
\centering
\caption{Best-fit parameters of ZTF20abbiixp / GRB~200524A for ISM like medium using the FS + RS model.}
\begin{tabular}{lccc}

\hline

Model & Prior range &  Best fit parameters \\
\hline\hline
& & &\\

\vspace{0.1cm}
$\theta_{\rm core}$ (rad)  & 0.087 - 1.040   & $0.273^{+0.022}_{-0.018}$  \\
\vspace{0.1cm}
$\log_{10} E_0$ (erg)   & 52 - 55      & $54.255^{+0.245}_{-0.204}$   \\
\vspace{0.1cm}
$\log_{10} n_{0}$(cm$^{-3}$) & (-2) - 3 & $2.822^{+0.112}_{-0.133}$ \\
\vspace{0.1cm}
$p$         & (2.01 - 2.40)                  & $2.04^{+0.007}_{-0.005}$   \\
\vspace{0.1cm}
$\log_{10} \epsilon_e$  & (-4) - (-0.6)      & $-2.960^{+0.194}_{-0.229}$  \\
\vspace{0.1cm}
$\log_{10} \epsilon_B$   &  (-3) - (-0.5)    & $-0.528^{+0.177}_{-0.269}$   \\
\vspace{0.1cm}
$p_{RS}$      & 2.01 - 2.40                & $2.083^{+0.010}_{-0.009}$   \\
\vspace{0.1cm}
$\log_{10} \epsilon_{e,RS}$ & (-6) - (-0.6)         & $-0.705^{+0.205}_{-0.268}$ \\
\vspace{0.1cm}
$\log_{10} \epsilon_{B,RS}$   & (-6) - (-0.5)       & $-5.456^{+0.152}_{-0.180}$   \\

$t_{\rm X}$        &  100 - 6000              & $4893.371^{+594.608}_{-1059.747}$           \\
& & &\\
\hline
\label{Tab:bestfit_fsrs}
\end{tabular}
\end{table}

\subsubsection{The afterglow of ZTF20abbiixp / GRB~200524A in comparison with other long soft GRB sample}

\begin{figure}
	\includegraphics[width=\columnwidth]{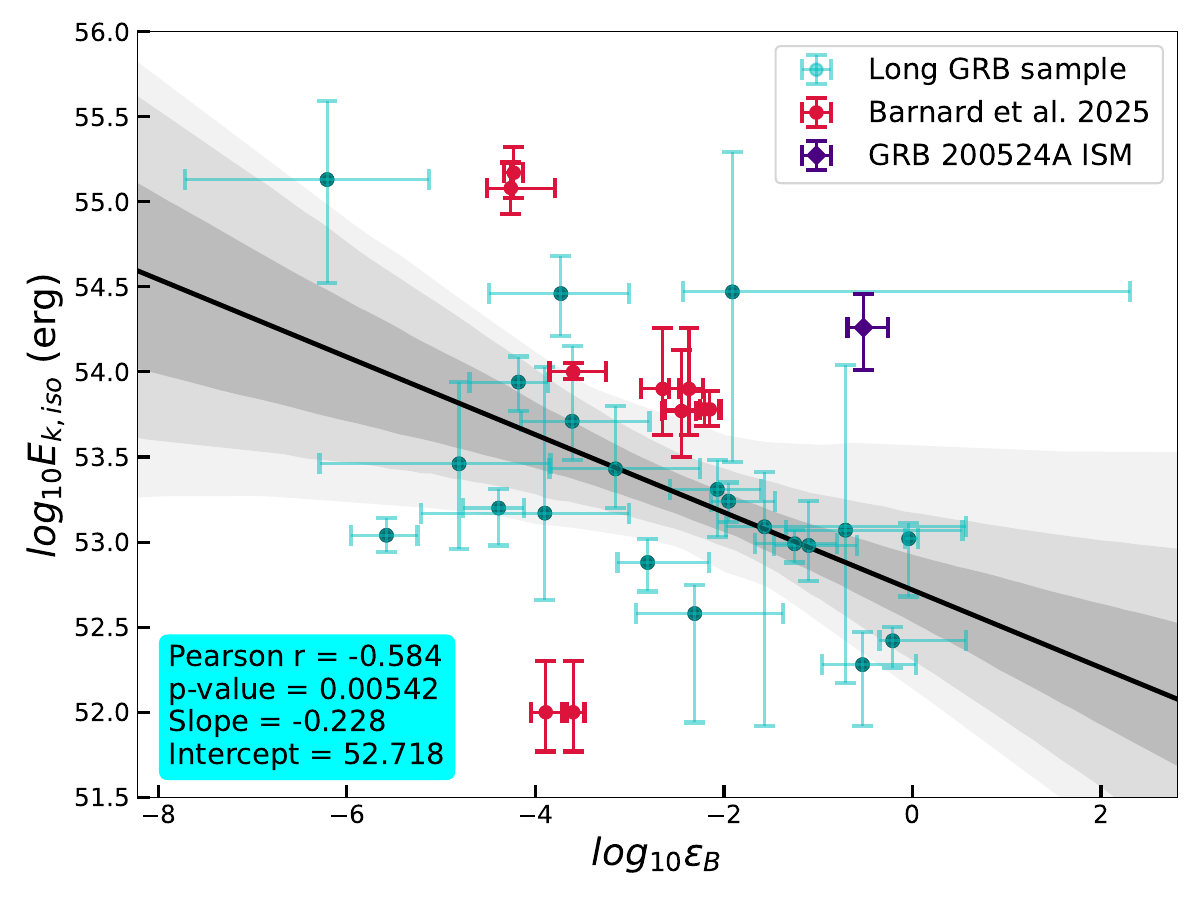}
    \caption{Correlation between ($E_{\rm K,iso}$), and ($\epsilon_{\rm B}$), inferred from GRB afterglow modeling. The cyan points represent the high-energetic long-GRB sample from \citet{2022MNRAS.511.2848A}, the red points show the sample from \citet{2025MNRAS.543.4218B}, and the purple point marks ZTF20abbiixp / GRB~200524A assuming an ISM environment. The black solid line shows the best-fitting linear model obtained only for the cyan long-GRB sample from \citet{2022MNRAS.511.2848A}, with the shaded regions representing the corresponding 1-$\sigma$, 2-$\sigma$, and 3-$\sigma$ uncertainty bands, respectively. The Pearson coefficient shown in the figure, (r = -0.584) with (p = 0.00542), was calculated using the cyan long-GRB sample from \citet{2022MNRAS.511.2848A}.}
    \label{fig:correlation}
\end{figure}

To compare the afterglow properties of ZTF20abbiixp / GRB~200524A in the context of a sample of high energetic GRBs ($E_{\rm k,iso} > 10^{53}$ erg), we compiled the sample of high energetic GRBs with known afterglow parameters from the literature \citep{2022MNRAS.511.2848A, 2010ApJ...711..641C, 2011ApJ...732...29C}. In order to extend the parameter space and completeness of the sample, we collected the broadband afterglow modeling results of VHE GRBs presented by \citet{2025MNRAS.543.4218B}. Fig.~\ref{fig:correlation} shows strong anti-correlation between isotropic equivalent kinetic energy ($E_{\rm k,iso}$), and the fraction of energy distributed to magnetic field ($\epsilon_B$) for high energetic long soft GRB sample reported by \citet{2022MNRAS.511.2848A}, with Pearson correlation coefficient r = -0.584 and a p-value of $5.4 \times 10^{-3}$. The strong anti-correlation clearly indicates that GRBs with high isotropic energy tend to occur in a poorly magnetized ambient medium.

The significance of the anti-correlation rapidly decreases with the inclusion of VHE GRBs. This is likely to be associated with the different radiative characteristics and shock microphysics inferred for VHE-detected bursts, many of which require alternative emission channels or distinct parameter regimes to explain their broadband SED \citep{2025MNRAS.543.4218B}. The dark and light shaded regions in Fig.~\ref{fig:correlation} indicate 1-$\sigma$ and 3-$\sigma$ significance levels from the mean. For consistency with previous studies, we considered only the FS values of $E_{\rm k,iso}$ and $\epsilon_B$ in this comparison. RS contributions were not considered, as they were generally excluded for the correlation in the literature \citep{2010ApJ...711..641C, 2011ApJ...732...29C, 2022MNRAS.511.2848A, 2025MNRAS.543.4218B}. ZTF20abbiixp / GRB~200524A lies beyond the 3-$sigma$ level. The position of ZTF20abbiixp / GRB~200524A is unique in the correlation.  

\section{Summary}
\label{discussion}

We carried out a comprehensive and detailed multi-wavelength study of ZTF20abbiixp / GRB~200524A, using an extensive data set spanning from $\gamma$-ray to radio data, observed with several space and ground based telescopes. The prompt emission timing analysis of ZTF20abbiixp / GRB~200524A from \textit{Fermi}-GBM, \textit{Integral} SPI-ACS, \textit{Wind}-Konus, and \textit{Astrosat}-CZTI revealed that ZTF20abbiixp / GRB~200524A is a multi-peaked long soft GRB with no precursor emission. Several overlapping pulses were observed in the logarithmic light curve, which deviates from the standard powerlaw emission. Time-integrated and time-resolved spectroscopy demonstrate that the Band function provides the best fit for most time slices of prompt emission, except for the peak and the final time bin. The evolution of the spectral parameters of the Band function reveals an intensity tracking pattern in which $E_{\rm p}$ evolves in tandem with the flux. In contrast, the $\alpha$ and $\beta$ do not follow the same trend. The $\alpha$ value mostly remains close to the synchrotron line of death for slow cooling ($\alpha \sim$ 2/3), except during the interval close to light curve maxima \citep{2002ApJ...581.1248P, 2013ApJS..208...21G}. This behavior suggests the synchrotron radiation is the dominant emission mechanism for the prompt emission of ZTF20abbiixp / GRB~200524A, while the peak of the light curve requires an additional component. Similarly, the maximum energetic photon (9.8 GeV) detected by \textit{Fermi}-LAT, shown in Fig.~\ref{fig:lat photon light curve}, exceeds the maximum energy expected from synchrotron emission, pointing to its non-synchrotron origin.

We further examined the prompt emission properties of ZTF20abbiixp/GRB~200524A, alongside those of other GRBs using standard correlations. Amati correlation, along with EH-$T_{90}$ correlation, supports the classification of ZTF20abbiixp / GRB~200524A as a long soft GRB. ZTF20abbiixp / GRB~200524A reveals almost zero spectral lag due to multiple overlapping pulses, which is not usually observed in the case of long soft GRBs \citep{2012MNRAS.419..614U, 2015MNRAS.446.1129B}. In the spectral lag - isotropic luminosity correlation, ZTF20abbiixp / GRB~200524A lies at the boundary between long soft and short hard GRBs with large uncertainties.

The optical light curves of ZTF20abbiixp / GRB~200524A exhibit a steep to shallow transition following the broken powerlaw with a transition at around 1 d since the GBM trigger. The broadband SED is well described by a broken powerlaw, with a break between the X-ray and optical bands. Following the afterglow closure relations by \citet{2018ApJ...866..162G} using the temporal and spectral indices, afterglow emission of ZTF20abbiixp / GRB~200524A does not identify the ambient medium. Spectroscopic observation with the Gemini-GMOS instrument revealed the prominent absorption lines corresponding to MgII, CaII, and CrII at a redshift of z = 1.256. The calculated value of redshift was further used for the multi-wavelength modeling. 

We model the broadband afterglow of ZTF20abbiixp / GRB~200524A using a custom-developed numerical afterglow framework that includes contributions from both FS and RS. This model assumes a top-hat jet propagating through a constant density ISM or a stratified wind-like profile. The multi-wavelength modeling of ZTF20abbiixp / GRB~200524A suggests the combination of FS and RS provides a better fit than the individual components. The FS shock alone fails to explain the early time optical data in the ISM medium. Inclusion of the RS emission in the ejecta solves the problem. The model incorporating the wind medium fails to explain the multi-wavelength datasets of ZTF20abbiixp / GRB~200524A. The modeling suggests ZTF20abbiixp / GRB~200524A is a high energetic GRB with isotropic kinetic energy of $E_{k,iso} \sim 1.79 \times 10^{54}$ erg occurring in the exceptionally high density ambient medium. Even after having a high number density, the wind model is unable to explain the data. Such a high number density of the ambient medium can be directly linked to the presence of dense gas in the vicinity of a star forming region \citep{2026NatAs.tmp...42T}. The inferred parameters from the broadband afterglow modeling favors dense ambient medium in which the FS efficiently amplifies magnetic fields ($\epsilon_{B,FS}$ = 0.3), whereas the ejecta crossing the RS remains weakly magnetized ($\epsilon_{B,RS}$ = 3.5 $\times \ 10^{-5}$). This clearly indicates that the magnetic field is produced predominantly at the external shock rather than inherited from the ejecta. The unique combination of high isotropic energy, dense ISM-like environment, and contrasting FS/RS magnetization values places this burst among an uncommon class of energetic long soft GRBs. These properties also explain why ZTF20abbiixp / GRB~200524A holds a unique position in the log($E_{\rm k,iso}$)-log($\epsilon_B$) plane.

In conclusion, ZTF20abbiixp / GRB~200524A provides an example of an high energetic long soft GRB for which the prompt and afterglow properties were interpreted through detailed multi-wavelength observations. The prompt-emission properties indicate a non-thermal origin where the peak energy follows the flux evolution, while the highest-energy LAT photon suggests that an additional high-energy emission component may also be required. The panchromatic afterglow observations and modeling favors an FS+RS origin in a dense ISM. This study therefore highlights the importance of rapid optical photometric and spectroscopic observations, coordinated multi-instrument follow-up, and broadband physical modeling to uncover the diversity of GRB environments and burst properties.

\section*{Acknowledgements}

This research has used the VizieR catalogue access tool, operated at CDS, Strasbourg, France (DOI: 10.26093/cds/vizier). The original description of the VizieR service was published in A$\&$AS 143, 23. This research was supported by the National Research Foundation
(NRF) of South Africa through a BRICS Multilateral Grant with number 150504 to SR. AG was supported by NRF through a Postdoctoral Fellowship.  Based on observations obtained at the 3.6m Devasthal Optical Telescope (DOT), which is a National Facility run and managed by Aryabhatta Research Institute of Observational Sciences (ARIES), an autonomous Institute under the Department of Science and Technology, Government of India. Dimple acknowledges support from STFC grant No.ST/Y002253/1. GROWTH India telescope is a 70-cm telescope with a 0.7 degree field of view, set up by the Indian Institute of Astrophysics and the Indian Institute of Technology Bombay with support from the Indo-US Science and Technology Forum (IUSSTF) and the Science and Engineering Research Board (SERB) of the Department of Science and Technology (DST), Government of India (https://sites.google.com/view/growthindia/). It is located at the Indian Astronomical Observatory (Hanle), operated by the Indian Institute of Astrophysics (IIA). GROWTH-India project is supported by SERB and administered by IUSSTF. Harsh Kumar thanks the LSSTC Data Science Fellowship Program, which is funded by LSSTC, NSF Cybertraining Grant \#1829740, the Brinson Foundation, and the Moore Foundation; his participation in the program has benefited this work. Based on observations collected at the Centro Astron\'omico Hispano-Alem\'an (CAHA) at Calar Alto, operated jointly by Junta de Andaluc\'ia and Consejo Superior de Investigaciones Cient\'ificas (IAA-CSIC). D. A. Kann acknowledges support from the Spanish research project RTI2018-098104-J-I00 (GRBPhot).  RG was sponsored by the National Aeronautics
and Space Administration (NASA) through a contract with ORAU. The views and conclusions contained in this document are those of the authors and should not be interpreted as representing the official policies, either expressed or implied, of the National Aeronautics and Space Administration (NASA) or the U.S. Government. The U.S. Government is authorized to reproduce and distribute reprints for Government purposes notwithstanding any copyright notation
herein




\bibliographystyle{mnras}
\bibliography{refag} 




\clearpage
\appendix

\section{Additional photometric data}
\label{app:photometry}

\begin{table*}

\centering
\caption{Photomeric observations of the afterglow of GRB~200524A}
\label{tab:optical_mag}
\setlength{\tabcolsep}{18pt}
 \begin{tabular}{c c c c c c} 
 \hline
UT Date & $\Delta$ t   &  Filter  &   Magnitude & Telescope &  Remarks\\
        &   (d)     &          &             &           &         \\
\hline
2020-05-25.571 & 1.360 & {\it white}  & $22.06\pm0.12$  & $1$  & \citep{UVOT200524A}\\
\hline
2020-05-24.861 & 0.650 & $B$  & $20.97\pm0.08$  & 2 & This Work \\ 
\hline
2020-05-24.883 & 0.671 & $V$  & $21.15\pm0.13$  & 2 & This Work \\
2020-05-26.917 & 2.705 & $V$  & $22.69\pm0.19$  & 2 & This Work \\
\hline
2020-05-24.797 & 0.585 &  $CR$ & $20.43\pm0.25$   & 3 & This Work \\ 
2020-05-25.291 & 1.080 & $CR$ & $21.2\pm0.20$  & 4 & \citep{Zheng200524A} \\
\hline
2020-05-24.549 & 0.338 & $R_{c}$ & $19.50\pm0.30$  & 5 & \citep{2020GCN.27813....1O} \\
2020-05-24.816 & 0.605 & $R_{c}$ & $20.48\pm0.05$  & 6 & This Work \\
2020-05-24.819 & 0.608 & $R_{c}$ & $20.52\pm0.05$  & 6 & This Work \\ 
2020-05-24.822 & 0.611 & $R_{c}$ & $20.55\pm0.05$  & 6  & This Work \\ 
2020-05-24.825 & 0.614 & $R_{c}$ & $20.60\pm0.08$  & 6  & This Work \\ 
2020-05-24.831 & 0.620 & $R_{c}$ & $20.59\pm0.09$  & 6  & This Work \\ 
2020-05-24.834 & 0.622 & $R_{c}$ & $20.60\pm0.09$  & 6 & This Work \\ 
2020-05-24.903 & 0.691 & $R_{c}$ & $20.91\pm0.24$  & 2 & This Work \\
2020-05-25.482 & 1.271 & $R_{c}$  & $> 18.70$  &  5 & \citep{2020GCN.27824....1H} \\
2020-05-26.937 & 2.725 & $R_{c}$ & $22.63\pm0.10$  & 2 & This Work \\
2020-05-27.909 & 3.697 & $R_{c}$ & $22.86\pm0.20$ & 7 & This Work\\
2020-05-28.910 & 4.702 & $R_{c}$ & $23.21\pm0.25$  & 2 & This Work \\
2020-05-28.929 & 4.718 & $R_{c}$ & $>22.6$  & 7 & This Work\\
2020-09-23.310 & 121.800 & $R_{c}$ & $>24.6$  & 6 & This Work\\
\hline
2020-05-24.549 & 0.338 & $I_{c}$ & $19.40\pm0.50$  & 5 & \citep{2020GCN.27813....1O} \\
2020-05-28.826 & 0.630 & $I_{c}$ & $20.32\pm0.22$  & 8 & This Work\\
2020-05-24.924 & 0.712 & $I_{c}$ & $20.38\pm0.16$  & 2 & This Work \\
2020-05-25.482 & 1.271 & $I_{c}$  & $> 18.00$  &  5 &  \citep{2020GCN.27824....1H} \\
2020-05-26.958 & 2.746 & $I_{c}$ & $21.79\pm0.13$  & 2 & This Work \\
\hline
2020-05-24.910 & 0.699 & $u^\prime$ & $21.57\pm0.24$ & 9 & This Work \\
\hline
2020-05-24.549 & 0.338 &  $g^\prime$ & $19.60\pm0.30$  & 5 & \citep{2020GCN.27813....1O} \\
2020-05-24.675 & 0.464 & $g^\prime$ & $20.80\pm0.17$ & 10 & This Work\\
2020-05-24.730 & 0.519 & $g^\prime$ & $21.16\pm0.47$ & 10 & This Work\\
2020-05-24.901 & 0.690 & $g^\prime$ & $21.10\pm0.07$  & 11 & This Work \\
2020-05-24.905 & 0.694 & $g^\prime$ & $21.14\pm0.07$  & 11 & This Work \\
2020-05-24.908 & 0.697 & $g^\prime$ & $21.267\pm0.06$ & 9 & This Work \\
2020-05-24.909 & 0.697 & $g^\prime$ & $21.24\pm0.08$  & 11 & This Work \\
2020-05-24.910 & 0.699 & $g^\prime$ & $21.49\pm0.05$  &  12 & This Work\\
2020-05-25.048 & 0.836 & $g^\prime$ & $21.61\pm0.05$  & 9 & This Work \\
2020-05-25.071 & 0.859 & $g^\prime$ & $21.54\pm0.10$ & 9 & This Work \\
2020-05-25.171 & 0.960 & $g^\prime$ & $21.58\pm0.16$ & 13 & This Work \\
2020-05-25.482 & 1.271 & $g^\prime$  & $> 18.62$  &  5 & \citep{2020GCN.27824....1H} \\
2020-05-25.789 & 1.578 & $g^\prime$ & $22.18\pm0.07$  &  12 & This Work \\ 
2020-05-25.814 & 1.603 & $g^\prime$ & $22.13\pm0.06$  &  12 & This Work \\
2020-05-25.999 & 1.787 & $g^\prime$ & $22.41\pm0.19$ & 9 & This Work \\
2020-05-26.213 & 2.001 & $g^\prime$ & $22.57\pm0.20$ & 13 & This Work \\
2020-05-26.904 & 2.693 & $g^\prime$ & $22.82\pm0.26$ & 9 & This Work \\
2020-05-28.018 & 3.807 & $g^\prime$ & $22.87\pm0.28$ & 9 & This Work \\
2020-06-23.344 & 30.133 & $g^\prime$ & $>26.76$ & 14 & This Work \\
\hline
2020-05-24.286 & 0.075 & $r^\prime$ & $17.10\pm0.01$ & 15 & This Work \\
2020-05-24.289 & 0.078 & $r^\prime$ & $17.34\pm0.01$ & 15 & This Work \\
2020-05-24.304 & 0.093 & $r^\prime$ & $17.59\pm0.01$ & 15 & This Work \\
2020-05-24.307 & 0.095 & $r^\prime$ & $17.63\pm0.01$ & 15 & This Work \\
2020-05-24.683 & 0.472 & $r^\prime$ & $20.57\pm0.14$ & 10 & This Work\\
2020-05-24.738 & 0.527 & $r^\prime$ & $20.59\pm0.22$ & 10 & This Work\\
2020-05-24.768 & 0.599 & $r^\prime$ & $20.80\pm0.08$  &  12 & This Work \\
2020-05-24.822 & 0.611 & $r^\prime$ & $20.90\pm0.07$  &  12  & This Work\\
2020-05-24.844 & 0.632 & $r^\prime$ & $21.01\pm0.16$ & 10 & This Work\\
2020-05-24.851 & 0.641 & $r^\prime$ & $21.04\pm0.18$ & 10 & This Work\\
2020-05-24.858 & 0.647 & $r^\prime$ & $21.01\pm0.17$ & 10 & This Work\\
2020-05-24.865 & 0.654 & $r^\prime$ & $20.91\pm0.15$ & 10 & This Work\\
2020-05-24.871 & 0.660 & $r^\prime$ & $21.00\pm0.18$ & 10 & This Work\\
2020-05-24.906 & 0.695 & $r^\prime$ & $20.99\pm0.07$ & 9 & This Work \\
2020-05-24.913 & 0.701 & $r^\prime$ & $20.99\pm0.07$  & 11 & This Work \\
2020-05-24.916 & 0.705 & $r^\prime$ & $21.06\pm0.07$  & 11 & This Work \\
2020-05-24.924 & 0.713 & $r^\prime$ & $21.19\pm0.07$  & 11 & This Work \\
2020-05-25.052 & 0.841 & $r^\prime$ & $21.36\pm0.05$  & 9  & This Work\\
2020-05-25.068 & 0.857 & $r^\prime$ & $21.31\pm0.09$ & 9 & This Work \\
\hline
  \end{tabular}
\end{table*}

\begin{table*}
\centering
\setlength{\tabcolsep}{18pt}
 \begin{tabular}{c c c c c c} 
 \hline
UT Date & $\Delta$ t   &  Filter  &   Magnitude & Telescope &  Remarks\\
        &   (d)     &          &             &           &         \\
\hline
2020-05-25.169 & 0.958 & $r^\prime$ & $21.59\pm0.21$ & 13 & This Work \\
2020-05-25.765 & 1.554 & $r^\prime$ & $22.09\pm0.13$  &  12  & This Work\\
2020-05-25.827 & 1.615 & $r^\prime$ & $22.11\pm0.12$  &  12  & This Work\\
2020-05-25.996 & 1.785 & $r^\prime$ & $22.08\pm0.14$ & 9 & This Work \\
2020-05-26.210 & 1.999 & $r^\prime$ & $22.24\pm0.23$ & 13 & This Work \\
2020-05-26.649 & 2.437 & $r^\prime$ & $22.56\pm0.15$  & 16 & This Work\\
2020-05-26.901 & 2.690 & $r^\prime$ & $22.32\pm0.19$ & 9 & This Work \\
2020-05-27.729 & 3.518 & $r^\prime$ & $22.94\pm0.19$  & 16 & This Work\\
2020-05-28.015 & 3.804 & $r^\prime$ & $22.81\pm0.28$ & 9 & This Work \\
2020-06-23.344 & 30.133 & $r^\prime$ & $>25.99$ & 14 & This Work \\
2021-02-05.359 & 257.148 & $r^\prime$ & $>25.4$  & 17 & This Work\\
\hline
2020-05-24.691 & 0.480 & $i^\prime$ & $20.60\pm0.29$ & 10 & This Work\\
2020-05-24.746 & 0.535 & $i^\prime$ & $>17.54$  & 10 & This Work\\
2020-05-24.900 & 0.689 & $i^\prime$ & $20.89\pm0.04$  &  12 & This Work \\
2020-05-24.904 & 0.693 & $i^\prime$ & $20.83\pm0.09$ & 9 & This Work \\
2020-05-24.926 & 0.715 & $i^\prime$ & $20.83\pm0.11$  & 11 & This Work \\
2020-05-24.929 & 0.718 & $i^\prime$ & $20.86\pm0.12$  & 11 & This Work \\
2020-05-24.931 & 0.720 & $i^\prime$ & $20.72\pm0.14$  & 11 & This Work \\
2020-05-25.057 & 0.846 & $i^\prime$ & $21.13\pm0.05$  & 9 & This Work \\
2020-05-25.066 & 0.855 & $i^\prime$ & $21.09\pm0.07$ & 9 & This Work \\
2020-05-25.172 & 0.961 & $i^\prime$ & $21.03\pm0.18$ &13 & This Work \\
2020-05-25.773 & 1.562 & $i^\prime$ & $21.96\pm0.14$  &  12 & This Work \\ 
2020-05-25.831 & 1.620 & $i^\prime$ & $22.24\pm0.11$  &  12 & This Work \\ 
2020-05-25.994 & 1.782 & $i^\prime$ & $22.07\pm0.27$ & 9 & This Work \\
\hline
2020-05-24.902 & 0.691 & $z^\prime$ & $20.47\pm0.15$ & 9 & This Work \\
2020-05-24.936 & 0.725 & $z^\prime$ & $20.72\pm0.10$  & 11 & This Work \\
2020-05-24.941 & 0.730 & $z^\prime$ & $21.04\pm0.28$  & 11 & This Work \\
2020-05-25.781 & 1.570 & $z^\prime$ & $21.95\pm0.11$  &  12 & This Work \\
2020-05-25.840 & 1.628 & $z^\prime$ & $22.03\pm0.25$  &  12 & This Work \\
2020-05-25.844 & 1.633 & $z^\prime$ & $22.24\pm0.16$  &  12 & This Work \\
2021-02-05.359 & 257.148 & $z^\prime$ & $>24.5$  & 17 & This Work\\

\hline

  \end{tabular}
\newline
\newline
\noindent
\noindent
\footnotesize{NOTE: $1$ - {\it Swift}-UVOT , $2$ - 2.2m CAHA/CAFOS, 3 - 0.36m Kitab-ISON RC-36, 4 - 0.76m KAIT, 5 - 0.5m MITSuME telescope, 6 - 2.6m CrAO Shajn Telescope, 7 - 1.0m SAO-RAS Zeiss Telescope, 8 - 1.3m Devasthal Faint Object Telescope, 9 - 2.0m Liverpool Telescope, 10 - 0.7m GROWTH-India Telescope, 11 - 0.8m OAJ T80, 12 - 3.6m Devasthal Optical Telescope/ADFOSC, 13 - 1.52m Palomar P60/SEDM, 14 - 10.0m Keck/LRIS, 15 - 1.22m Palomar P48 Schmidt/ZTF, 16 - 1.5m Assy-Turgen AZT-20 Telescope, 17 - 8.4m Large Binocular Telescope. $\Delta t$ is observer-frame time relative to the GBM trigger time $T_0$ given in Section \ref{GBM}}

\end{table*}

\section{Comparison with other energetic GRBs}
\label{app:comparison}

\begin{table}
\caption{Comparison of afterglow model parameters of ZTF20abbiixp / GRB~200524A w.r.t other high energetic GRBs}  
\label{tab:comparison}
 \begin{tabular}{cccc}
  \hline
GRB name & $log_{10}E_{k,iso}$  &  $log_{10}n_0$  & $log_{10}\epsilon_B$   	\\
\hline \hline

\addlinespace[0.25cm]
970508  &   $53.20_{-0.21}^{+0.22}$      & $2.18_{-0.14}^{+0.11}$   &  $-4.39_{-0.27}^{+0.38}$    \\
\addlinespace[0.15cm]
990510     & $52.99_{-0.08}^{+0.11}$      & $-0.95_{-0.35}^{+0.15}$   &  $-1.25_{-0.45}^{+0.42}$    \\
\addlinespace[0.15cm]
991208    & $54.64_{-0.41}^{+0.10}$      & $-0.60_{-0.11}^{+0.19}$   &  $-0.07_{-0.16}^{+0.07}$  \\
\addlinespace[0.15cm]
991216    & $53.71_{-0.44}^{+0.23}$     &  $1.28_{-0.52}^{+0.23}$   &  $-3.61_{-0.82}^{+0.54}$    \\
\addlinespace[0.15cm]
000418   & $54.47_{-0.82}^{+1.00}$      &  $1.11_{-1.12}^{+0.70}$   &  $-1.91_{-4.22}^{+0.52}$    \\
\addlinespace[0.15cm]
000926 & $55.13_{-0.46}^{+0.61}$        &  $1.99_{-0.39}^{+0.79}$   &  $-6.21_{-1.08}^{+1.51}$   \\
\addlinespace[0.15cm]
010222  &   $53.94_{-0.15}^{+0.17}$    &   $-2.32_{-0.17}^{+0.19}$   & $-4.18_{-0.31}^{+0.52}$     \\
\addlinespace[0.15cm]
030329 &    $53.04_{-0.10}^{+0.10}$    &   $2.59_{-0.18}^{+0.20}$   &  $-5.58_{-0.33}^{+0.38}$     \\
\addlinespace[0.15cm]
050820A &   $54.61_{-0.02}^{+0.03}$    &   $-3.40_{-0.13}^{+0.10}$   &  $-1.95_{-0.49}^{+0.19}$     \\
\addlinespace[0.15cm]
050904  &   $53.31_{-0.17}^{+0.28}$    &  $1.05_{-0.87}^{+0.25}$    &  $-2.07_{-0.46}^{+0.50}$     \\
\addlinespace[0.15cm]
090328  &   $53.17_{-0.86}^{+0.51}$    &  $1.66_{-0.79}^{+0.70}$    &  $-3.90_{-0.89}^{+1.31}$     \\
\addlinespace[0.15cm]
090423  &   $53.46_{-0.48}^{+0.50}$    &  $2.01_{-0.62}^{+0.75}$    &  $-4.81_{-0.97}^{+1.48}$     \\
\addlinespace[0.15cm]
090902B &   $53.74_{-0.02}^{+0.05}$    &  $-3.25_{-0.05}^{+0.07}$    &  $-3.15_{-0.91}^{+0.70}$     \\
\addlinespace[0.15cm]
090926A &   $53.09_{-0.32}^{+1.17}$    &  $0.19_{-0.26}^{+2.07}$    &  $-1.57_{-0.23}^{+0.29}$    \\
\addlinespace[0.15cm]
120521C &   $52.98_{-0.26}^{+0.21}$    &  $-0.47_{-0.29}^{+0.41}$   &  $-1.10_{-0.51}^{+0.37}$    \\
\addlinespace[0.15cm]
130427A &   $52.69_{-0.07}^{+0.05}$    &  $-2.53_{-0.08}^{+0.10}$   &  $0.20_{-0.06}^{+0.13}$    \\
\addlinespace[0.15cm]
130702A &   $53.07_{-0.97}^{+0.90}$    &  $-0.90_{-0.45}^{+1.42}$   &  $-0.71_{-1.24}^{+0.63}$     \\
\addlinespace[0.15cm]
130907A &   $53.02_{-0.09}^{+0.34}$    &  $-1.26_{-0.11}^{+0.12}$   &  $-0.04_{-0.10}^{+0.04}$     \\
\addlinespace[0.15cm]
140904A &   $54.46_{-0.42}^{+0.25}$    &  $1.96_{-0.35}^{+0.57}$    &  $-3.73_{-0.72}^{+0.36}$     \\
\addlinespace[0.15cm]
\textbf{200524A} & $54.255^{+0.245}_{-0.204}$    &   $2.822^{+0.112}_{-0.133}$  &  $-0.528^{+0.177}_{-0.269}$     \\
\addlinespace[0.15cm]
\hline
\end{tabular}

\end{table}

\section{High-energy light curves}
\label{app:highenergy}

\begin{figure*}
    \centering
    \begin{minipage}{0.48\textwidth}
        \centering
        \includegraphics[width=\linewidth]{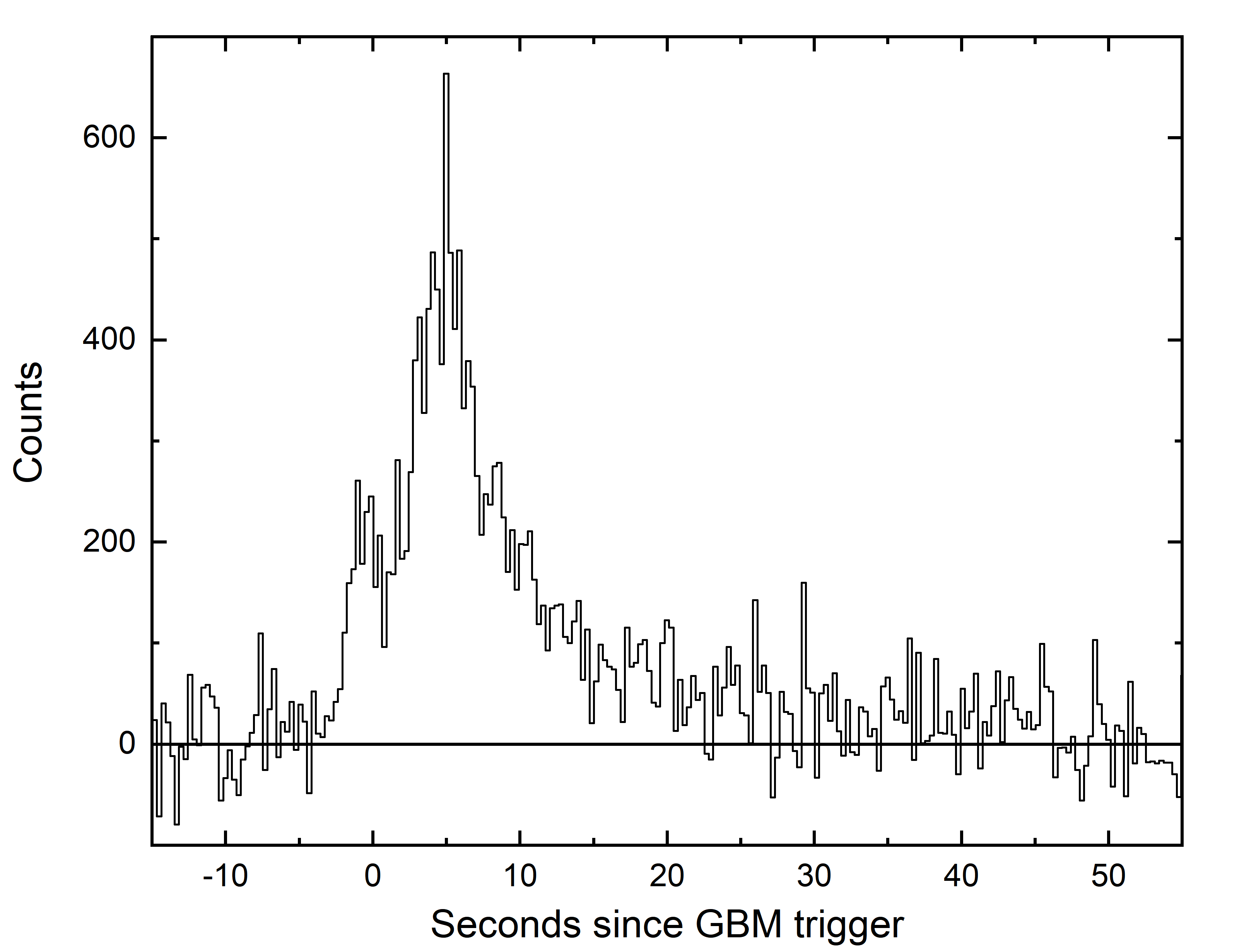}
        \caption{Light curve of ZTF20abbiixp/GRB~200524A with a time resolution of 0.3 s above 80 keV, based on \textit{INTEGRAL}-SPI-ACS data.}
        \label{fig:acs_lc}
    \end{minipage}
    \hfill
    \begin{minipage}{0.48\textwidth}
        \centering
        \includegraphics[width=\linewidth]{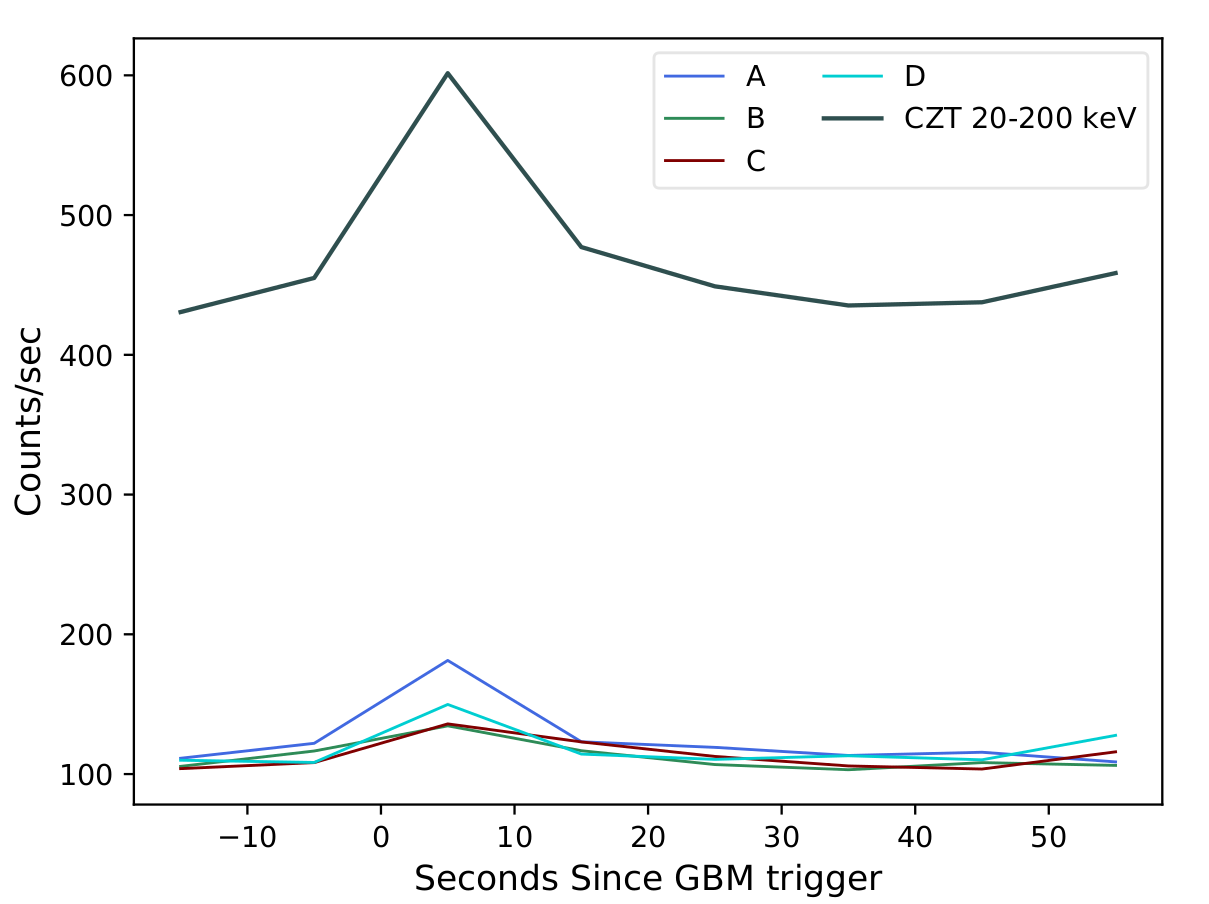}
        \caption{Composite and quadrant-wise light curves of ZTF20abbiixp/GRB~200524A in the 20--200 keV range, based on \textit{AstroSat}-CZTI data.}
        \label{fig:astrosat_lc}
    \end{minipage}
\end{figure*}

\section{Corner plot}
\label{app:corner}

\begin{figure*}
    \centering
    \includegraphics[width=\textwidth]{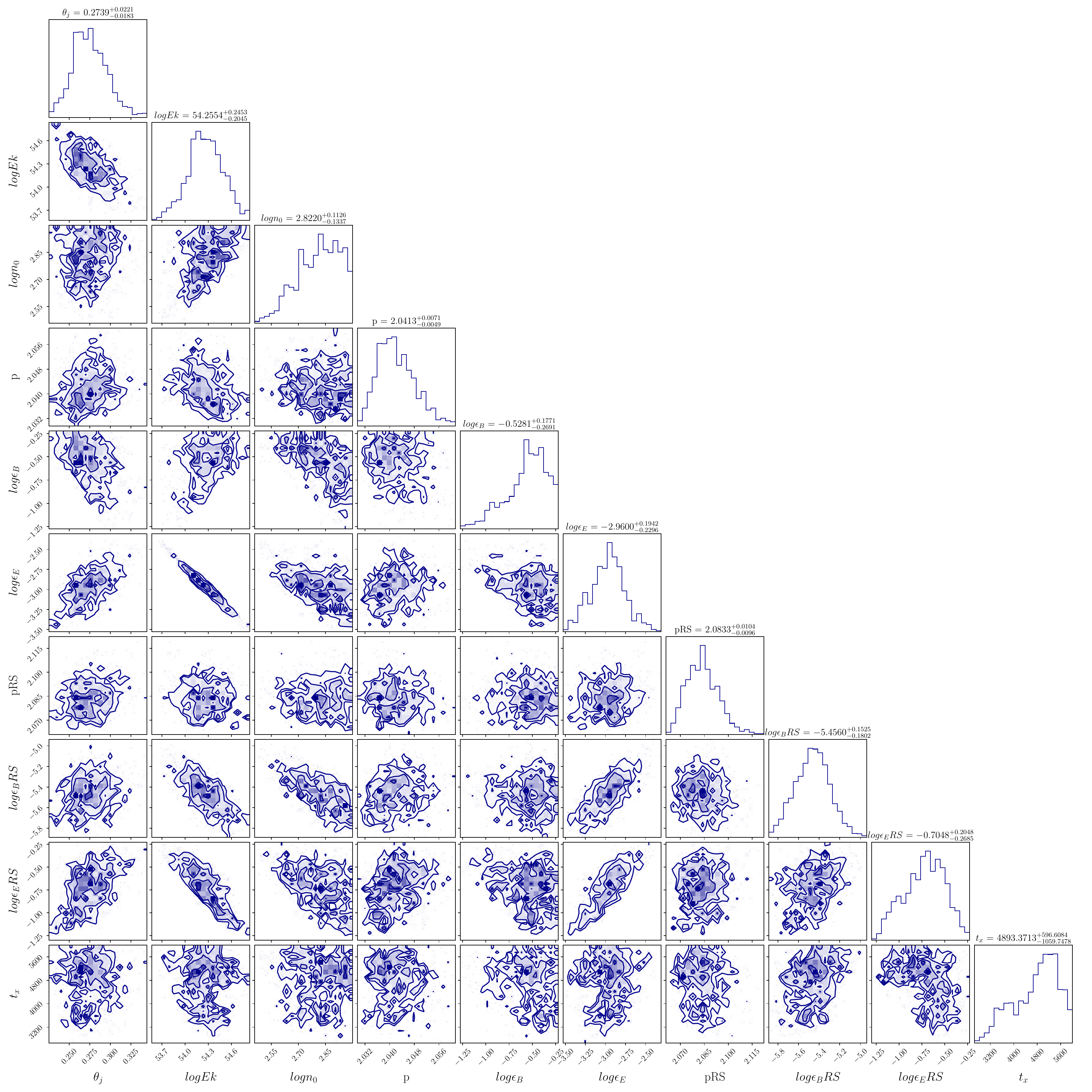}
    \caption{Corner plot for ZTF20abbiixp/GRB~200524A showing the posterior distributions and correlations of the afterglow model parameters.}
    \label{fig:corner}
\end{figure*}

$^{1}$Centre for Astro-Particle Physics (CAPP), Department of Physics, University of Johannesburg, PO Box 524, Auckland Park 2006, South Africa\\
$^{2}$School of Physics and Astronomy, University of Birmingham, Edgbaston, Birmingham, B15 2TT, UK.\\
$^{3}$Institute for Gravitational Wave Astronomy, University of Birmingham, Birmingham, B15 2TT, UK.\\
$^{4}$Aryabhatta Research Institute of observational sciences, Manora Peak, Nainital 263 001, India\\
$^{5}$Space Research Institute, Russian Academy of Sciences, Moscow, 117997, Russia\\
$^{6}$Cahill Center for Astrophysics, California Institute of Technology, MC 249-17, 1200 E California Boulevard, Pasadena, CA, 91125, USA\\
$^{7}$Instituto de Astrof\'isica de Andaluc\'ia (IAA-CSIC), Glorieta de la Astronom\'ia, s/n, E-18008, Granada, Spain\\
$^{8}$ Faculty of Physics, Higher School of Economics, Moscow 101000, Russia\\
$^{9}$ School of Physics \& Astronomy, Monash University, Clayton VIC 3800, Australia \\
$^{10}$ INAF, Osservatorio Astronomico di Capodimonte, Salita Moiariello 16, I-80131, Napoli, Italy\\
$^{11}$ DARK, Niels Bohr Institute, University of Copenhagen, Jagtvej 155A, 2200, Copenhagen N, Denmark\\
$^{12}$ Physics Department, Indian Institute of Technology Bombay, Powai, 400 076, India \\
$^{13}$ Center for Astrophysics | Harvard \& Smithsonian, 60 Garden Street, Cambridge, MA 02138-1516, USA \\
$^{14}$ Universit\'e de la C\^ote d'Azur, Observatoire de la C\^ote d'Azur, CNRS, Artemis, 06304 Nice, France\\
$^{15}$ Aix Marseille Univ, CNRS, LAM Marseille, France\\
$^{16}$ INAF - Osservatorio di Astrofisica e Scienza dello Spazio, via Piero Gobetti 93/3, 40129 Bologna, Italy \\
$^{17}$ Indian Institute of Astrophysics, 2nd Block 100 Feet Rd, Koramangala Bangalore, 560 034, India\\
$^{18}$ Department of Physics, Ashoka University, Sonipat, Haryana-131029, India\\
$^{19}$ School of Studies in Physics and Astrophysics, Pandit Ravishankar Shukla University, Raipur, Chattisgarh 492010, India\\
$^{20}$Indian Institute of Space Science and Technology, Trivandrum 695547, Kerala, India\\
$^{21}$Astrophysics Science Division, NASA Goddard Space Flight Center,
Mail Code 661, Greenbelt, MD 20771, USA\\ 
$^{22}$ NASA Postdoctoral Program Fellow\\
$^{23}$Space Astronomy Group, ISITE Campus, U. R. Rao Satellite Centre, Bangalore, 560037, India.\\
$^{24}$ Department of Physics, Royal Holloway, University of London, Egham, TW20 0EX, UK\\
$^{25}$  South-Western Institute for Astronomy Research, Yunnan University, Kunming, Yunnan 650504, People’s Republic of China \\
$^{26}$ Yunnan Key Laboratory of Survey Science, Yunnan University, Kunming, Yunnan 650500, People’s Republic of China \\
$^{27}$ Special Astrophysical Observatory, Russian Academy of Sciences, Nizhnij Arkhyz, 369167, Russia \\
$^{28}$ Indian Institute of Technology, Kanpur, 208016, India\\
$^{29}$ Tata Institute of Fundamental Research, Homi Bhabha Road, Mumbai, India 400 005 \\
$^{30}$ Crimean Astrophysical Observatory, Russian Academy of Sciences,  Nauchny 298409, Crimea\\
$^{31}$ Petrozavodsk State University, Petrozavodsk, 185910, Russia \\
$^{32}$ Keldysh Institute of Applied Mathematics, Moscow, Russia\\
$^{33}$ Fesenkov Astrophysical Institute, Almaty, 050020, Kazakhstan \\
$^{34}$ Ulugh Beg Astronomical Institute, Uzbekistan Academy of Sciences, Tashkent 100052, Uzbekistan \\
$^{35}$ Samarkand State University, 15, University boulevard, Samarkand 140104, Uzbekistan\\

\bsp	
\label{lastpage}
\end{document}